\documentclass[format=acmsmall, review=false]{acmart}
\pdfoutput=1
\usepackage{acm-ec-23}
\usepackage{booktabs} 
\usepackage[ruled]{algorithm2e} 

\SetAlFnt{\small}
\SetAlCapFnt{\small}
\SetAlCapNameFnt{\small}
\SetAlCapHSkip{0pt}
\IncMargin{-\parindent}

\usepackage{amsmath}
\usepackage{graphicx}
\usepackage{float}
\usepackage{caption}
\usepackage{amsthm}
\usepackage{color}
\definecolor{green}{rgb}{0,0.5977,0}

\newcommand{\rr}{\mathbb R}

\newcommand{\zz}{\mathbb Z}

\newcommand{\suchthat}{\ | \ }

\newcommand{\genseq}[3]{{#1}_1 {#3} {#1}_2 {#3} \dots {#3} {#1}_{#2}}
\newcommand{\seq}[2]{\genseq{#1}{#2}{,}}

\newcommand{\twocases}[4]{\begin{cases} #2 & #1 \\ #4 & #3 \end{cases}}

\newcommand{\txt}[1]{\text{#1}}

\newcommand{\stext}[1]{\ \ \ \ \ \text{(#1)}}

\newcommand{\ipncm}[3]{\begin{figure}\begin{center}\includegraphics[scale = {#1}]{Media/#2.pdf}\caption{#3}\end{center}\end{figure}}

\newcommand{\igncm}[3]{\begin{figure}\begin{center}\includegraphics[scale = {#1}]{Media/#2.png}\caption{#3}\end{center}\end{figure}}

\makeatletter 
\g@addto@macro{\@algocf@init}{\SetKwInOut{Parameter}{Parameters}} 
\makeatother

\theoremstyle{plain}
\newtheorem{theorem}{Theorem}
\newtheorem{lemma}[theorem]{Lemma}

\usepackage{xcolor}
\usepackage{caption}
\usepackage{subcaption}

\definecolor{niceColorRed}{HTML}{b22222}
\definecolor{niceColorGreen}{HTML}{2ca02c}
\definecolor{niceColorBlue}{HTML}{4169e1}
\definecolor{niceColorBrown}{HTML}{CC6600}
\definecolor{niceColorPurple}{HTML}{7156a8}
\definecolor{green}{HTML}{009900}
\newcommand{\emdash}{\,---\,}

\setcitestyle{acmnumeric}

\title{You Can Have Your Cake and Redistrict It Too}

\author{Gerdus Benad\`e}
\affiliation{\institution{Boston University}}
\author{Ariel D. Procaccia}
\author{Jamie Tucker-Foltz}
\affiliation{\institution{Harvard University}}

\begin{abstract}
The design of algorithms for political redistricting generally takes one of two approaches: optimize an objective such as compactness or, drawing on fair division, construct a protocol whose outcomes guarantee partisan fairness. We aim to have the best of both worlds by optimizing an objective subject to a binary fairness constraint. As the fairness constraint we adopt the geometric target, which requires the number of seats won by each party to be at least the average (rounded down) of its outcomes under the worst and best partitions of the state.

To study the feasibility of this approach, we introduce a new model of redistricting that closely mirrors the classic model of cake-cutting. This model has two innovative features. First, in any part of the state there is an underlying ``density'' of voters with political leanings toward any given party, making it impossible to finely separate voters for different parties into different districts. This captures a realistic constraint that previously existing theoretical models of redistricting tend to ignore. Second, parties may disagree on the distribution of voters\,---\,whether by genuine disagreement or attempted strategic behavior. In the absence of a ``ground truth'' distribution, a redistricting algorithm must therefore aim to simultaneously be fair to each party with respect to its own reported data. Our main theoretical result is that, surprisingly, the geometric target is always feasible with respect to arbitrarily diverging data sets on how voters are distributed.

Any standard for fairness is only useful if it can be readily satisfied in practice. Our empirical results, which use real election data and maps of six US states, demonstrate that the geometric target is always feasible, and that imposing it as a fairness constraint comes at almost no cost to three well-studied optimization objectives.
\end{abstract}

\begin{document}

 
\maketitle

\section{Introduction}\label{secIntro}

To be elected to the U.S.~House of Representatives, a candidate  must win a plurality election in their district. These districts are redrawn every decade based on the most recent census; the composition and creation of districts are governed by both federal and state laws. At the federal level, the Voting Rights Act requires that districts be drawn to allow minority groups to fully participate in the democratic process. Locally, many states expect districts to be contiguous and several require districts to be compact and respect ``communities of interest.'' 

These guidelines, however, are often open to interpretation. For example, only six states specify a  metric by which compactness is measured; elsewhere the determination of whether or not a district is compact is based on rules of thumb. \emph{Gerrymandering}\footnote{The term dates back to then-Governor of Massachusetts Elbridge Gerry's 1812 approval of a salamander-shaped district that was thought to aid his Democratic-Republican Party.} is the process of exploiting this flexibility by carefully drawing district boundaries for political gain, for example to protect an  incumbent or to benefit (or suppress) a specific class, race or political party. It is widely recognized as a distortion of the democratic system; in recent years, mathematicians and computer scientists have mobilized to help address this issue~\cite{Duchin18b}.

One place where scientists can contribute is the design of rigorous methods for drawing electoral district maps, which we refer to as \emph{partitions}. This problem is often approached from an optimization perspective \cite{GN70,MJN98,Shirabe05,Shirabe09,OH17}, which involves setting an objective\,---\,such as compactness, or the number of ``competitive'' districts\,---\,and finding the optimal partition satisfying the legal constraints (e.g., contiguity, population equality). However, optimization-based approaches do not necessarily lead to \emph{fair} outcomes that would be acceptable to both major political parties.

\medskip
\noindent\textbf{Our approach.}
To address the shortcomings of the pure optimization-based approach, we propose to combine it with ideas from \emph{fair division}~\cite{BT96,Moul03} in a way that ideally enjoys the best of both worlds. On a high level, we wish to enforce an intuitive yet rigorous notion of fairness that is also binary, in the sense that it either is or is not satisfied\,---\,there is no question of degree. One key advantage of such a notion is that it would allow a simple \emph{explanation} of why a partition satisfying it is fair~\cite{Pro19}. Among all valid partitions that satisfy the fairness notion, we find one that optimizes a given objective function. This approach\,---\,optimizing an objective function subject to a binary fairness guarantee\,---\,is akin to recent practical success stories in fair division, such as a rent division algorithm~\cite{GMPZ17} that has been used to solve tens of thousands of real-world instances.

A key question, of course, is which fairness notion to use. One natural (albeit flawed) answer is \emph{proportionality}: the number of seats won by each party should be proportional to its statewide support. Unfortunately proportionality is not a feasible standard~\cite{Nag17}. For example, the Republican party won roughly 32\% of the Massachusetts statewide vote in the 2016 presidential election. Proportionality suggests that Republicans should win three (roughly 32\%) of the state's nine congressional seats. However, this is impossible: there is no partition of the state into nine districts that complies with Massachusetts' redistricting laws under which the Republican party wins any congressional seats based on this election data \cite{DGH+18}, as the distribution of Republican-leaning voters across the state is rather homogeneous. This is not necessarily disturbing in and of itself;  Supreme Court rulings ``clearly foreclose any claim that the Constitution requires proportional representation'' \cite{DvB}.

Instead, we employ the \emph{geometric target} criterion of \citet{LS14}. To motivate it from our own viewpoint, imagine a procedure in which a fair coin is flipped, and whichever party wins the coin flip is given absolute power to redistrict the state as they wish (subject to the relevant laws regarding contiguity, population equality \emph{etc.}). This  procedure would lead to extremely partisan partitions \emph{ex post}, that is, after the coin is flipped. However, it is certainly impartial \emph{ex ante} (before the coin is flipped), as every party is equally likely to suffer or benefit from it. The geometric target distills the essence of what makes this procedure fair, while avoiding its extreme partisan outcomes: each party must win the expected number of districts it would win under the above procedure, rounded down. In other words, the geometric target is the average, rounded down, of the maximum number of districts the party would win under any partition that satisfies the legal constraints, and the minimum number of districts the party would win under any such partition. We say that a partition is a \emph{GT partition} if the number of districts each party wins is at least its geometric target.\footnote{Rounding is necessary, since it is impossible to guarantee that two parties each win, say, at least 4.5 districts out of nine.}

For example, take the 2011 redistricting of Pennsylvania, which the state's Supreme Court ultimately struck down as unconstitutional and replaced with a remedial plan \cite{LWVvP}. The political poll aggregation website FiveThirtyEight published an ``Atlas of Redistricting''\footnote{See: \url{https://projects.fivethirtyeight.com/redistricting-maps/}} in which they study redistricting across the United States. Part of this effort involved constructing gerrymandered partitions that favor either of the major political parties. Taking these partitions as the most extreme outcomes and evaluating on the presidential election data from 2016, we find that the pro-Democratic map leads to nine Democratic congressional seats (out of 18) while the pro-Republican map leads to five Democratic seats. Based on this, the geometric target of the Democratic party (the average of their extreme outcomes) is seven, compared to the five won under the 2011 plan.

A possible objection is that the guarantee given by the geometric target depends on the underlying election data, which can be another source of contention\emdash what happens if the two parties disagree on which dataset should be used to evaluate targets? One of our conceptual contributions is that we explicitly allow the geometric targets of the two parties to be computed with respect to two different datasets. Thus, no matter whether the discrepancies arise from genuine informational disparities or deliberate attempts to achieve a more desirable outcome by manipulating data, any honest party should be satisfied by the final redistricting outcome. This is analogous to the guarantees of cake-cutting protocols: players may disagree over what parts of the cake are valuable, and the protocol must nevertheless find an allocation that is fair for all players according to their respective valuation functions.

As intuitively appealing as this extension of the geometric target is, however, it would not be useful if it cannot be enforced\emdash and so far there has been scant evidence that it can. Even if it can be enforced, it could conceivably restrict the space of feasible partitions to the point of significantly harming standard optimization objectives like compactness. This motivates our research questions:
\begin{quote}
	\emph{Do GT partitions exist in theory and are they feasible in practice? If so, is the geometric target compatible with standard optimization objectives?}
\end{quote}
The validity of our proposed approach hinges on the answers to both questions being positive. (Spoiler alert: they are.)

\medskip
\noindent\textbf{Our results.}
To develop a theoretical understanding of the existence of GT partitions, in Section \ref{secModel} we introduce a novel model of redistricting which we call the \emph{state-cutting} model. There are no inherent ``geometric'' constraints on what districts are allowed; instead, we abstract from real life the key challenge that geometry often presents: that the supporters of the two parties cannot be arbitrarily divided between districts. Thus, in the state-cutting model, every part of the state has an underlying density of support for each party. As the name suggests, to capture these density constraints we draw on ideas from the classic cake-cutting model~\cite{BT96,RW98,Pro13}, where densities are defined on the unit interval $[0, 1]$. Under this interpretation, we conceptualize redistricting as the act of partitioning $[0, 1]$ into \emph{districts}, each of which is a finite union of closed intervals (mirroring the typical assumption about pieces of cake).

In Section \ref{secProofs} we present our main theoretical result (Theorem \ref{thmUnboundedExistence}), that GT partitions always exist in the state-cutting model, even when the geometric targets of the two parties are computed with respect to two different pairs of density functions (corresponding to two different datasets). Our result is proved via a novel ``cut-and-choose'' protocol whereby one party divides a strategically critical subset of the interval into two equal pieces and the other party decides which party controls redistricting over which piece.

In Section \ref{secPractice} we empirically assess the quality of GT partitions in terms of the optimization objectives of compactness, efficiency gap and the number of competitive districts in six U.S.\txt{} states. We find that restricting our search to GT partitions rarely leads to a significant decrease in any of the three objectives, regardless of whether or not parties agree on the voter distribution. We conclude that the price of enforcing geometric targets as a notion of fairness is extremely low.

\medskip
\noindent\textbf{Related work.}
The connection between redistricting and fair division has inspired several papers that put forward interactive protocols by which the parties take turns splitting the state and choosing pieces~\cite{LRY09,LS14,PPY17,DGKO18,Brams19,ThresholdElection}. Of those, our work is most closely related to that of \citet{LS14}, who introduced the geometric target. They analyze the \emph{LRY protocol} of \citet{LRY09}, in which a neutral administrator presents both parties with a sequence of bipartitions $(L_1, R_1), (L_2, R_2), \dots, (L_{m - 1}, R_{m - 1})$ of the state into two pieces, with each $L_i \subseteq L_{i + 1}$. For each bipartition, both parties are asked whether they would rather redistrict $L_i$ into $i$ districts or $R_i$ into $m - i$ districts, with the other party redistricting the other side. If a point of agreement cannot be found, then there must be a specific $i$ at which both parties would prefer redistricting $R_i$ to $L_i$, but prefer redistricting $L_{i + 1}$ to $R_{i + 1}$, so randomness is used to determine whether to use partition $i$ or $i + 1$, and which party controls which piece. Landau and Su observe that, \emph{if} the feasible set of electoral maps is constrained to respect a given bipartition, then \emph{at least one} of the two options the parties are asked to choose between must meet their geometric target. However, this does not imply that the final outcome selected by the LRY protocol satisfies the geometric target itself, even for the party whose preferred choice was selected. Landau and Su acknowledge this shortcoming and informally argue that it is unlikely to cause serious problems in practice, appealing to the random elements of the protocol and the neutrality of the administrator.

\citet{DGKO18} provide a more rigorous treatment of the theoretical guarantees of the LRY protocol, showing that, in the absence of any geometric constraints, both parties are guaranteed to win at least two seats fewer than their geometric targets. However, under a simple grid-based model with a moderate, plausible compactness constraint, they show that the number of districts won by a party can be arbitrarily far from the geometric target. To the best of our knowledge, our paper presents the first protocol that provably satisfies the geometric targets of both parties under a nontrivial model.

We believe that our framework of considering multiple datasets presents a methodological innovation that such prior works have been lacking. It is often unclear exactly what problems gamified redistricting protocols are meant to be solving, since their only theoretical guarantees hold with respect to ground-truth data about where voters are distributed. One might reasonably ask why, in a setting where it is possible to objectively evaluate the fairness of any outcome, is an interactive protocol needed at all? We hope our approach, which uses a protocol to establish the existence of a fair outcome under diverging viewpoints of what is fair, will prove useful in setting the literature on fair division and redistricting on a robust foundation.

Beyond the fair-division viewpoint,  \emph{partisan symmetry} \cite{GK07,ND78, Jack94} and the \emph{efficiency gap} \cite{SM15} are alternative notions aimed at measuring how partisan a proposed plan is. Partisan symmetry ensures anonymity by requiring that parties are treated identically in the sense that each party would win the same number of seats as the other when they receive any particular fraction of the vote. To determine whether a partition in which one party wins 65\% of the seats with 53\% of the votes is impartial according to partisan symmetry, we must evaluate the number of seats the other party would have won had they received 53\% of the votes; indeed, this comparison must be done for the entire spectrum of potential outcomes. These hypothetical outcomes are typically generated by starting from a real election outcome (or a combination of several) and applying uniform \cite{Butler52} or approximately uniform swings \cite{King89,GK94} to model changes in voters' political preferences. Practically, uniform swings do not allow for the types of changes in voter preferences that occur in reality, and requiring partisan symmetry under more general models of electoral systems can be infeasible. 
The efficiency gap measures the net difference in the fraction of each party's wasted votes\emdash every vote cast for the minority in a district is deemed to have been wasted, as are all votes for the majority above the threshold required to win the district. Classic gerrymandering techniques like packing (concentrating a party's supporters in one district) and cracking (splitting a party's supporters into minorities in across many districts) lead to large efficiency gaps. A maximum efficiency gap threshold of 8\% has been proposed, although there are instances where this is impossible to attain.

On the optimization side, recent work has studied computational methods for redistricting from the perspective that there is an inherent trade-off between fairness and compactness \cite{Tradeoffs, SwamyJacobson, Fairmandering, HouseOnThePrairie}. Under cardinal measures of fairness such as proportionality or the efficiency gap, there is a ``Pareto-frontier'' of optimal partitions, at which improving fairness comes at a cost to compactness, and vice versa. Our approach is fundamentally different because our fairness condition is a binary constraint. Thus, our frontier necessarily has only two points: the most compact partition, and the most compact partition satisfying the geometric targets of both parties. In contrast to the recent work of \citet{Tradeoffs}, we find that the trade-off is not significant, which is a testament to the robustness and usefulness of the geometric target as a fairness requirement.

Further afield, the classical cake-cutting problem and its close relatives have received significant attention in computer science in general~\cite{Pro13} and in theoretical computer science in particular~\cite{EP06,AM16,DFHY18,ABR19}. A strength of our paper is that it provides a fundamentally different view of, and a new application domain for, this well-studied problem.

\section{The State-Cutting Model}\label{secModel}
Heterogeneous support throughout the state is captured by intregrable density functions over $[0, 1]$. A \emph{district} is a subset of $[0, 1]$ that can be expressed as a finite union of closed intervals. An instance of the \emph{state-cutting problem} is specified by a target number of districts $m \in \zz_{\geq 1}$, a set of $n$ parties $N$, and a set of $n^2$ \emph{voter distribution functions} $\{v_i^j \suchthat i, j \in N\}$ giving the measure of support for party $j$ according to party $i$ over any district. (We only concern ourselves with the case where $N = \{1, 2\}$ in this paper.) We assume that each $v_i^j$ is consistent with a measurable density function $f_i^j: [0, 1] \to [0, 1]$, where, for any district $D$,
$$v_i^j(D) = \int_{D}f_i^j(x) dx.$$
We additionally assume that the population density has been normalized so that, for any $x \in [0, 1]$ and $i \in N$,
$$\sum_{j \in N} f_i^j(x) = 1.$$
This implies that, for any district $D$ and party $i$,
$$\sum_{j \in N} v_i^j(D) = \mu(D),$$
where $\mu(D)$ is the measure of $D$. Figure \ref{figTheoreticalExample1} begins a hypothetical running example instance of the state-cutting problem.

\ipncm{0.5}{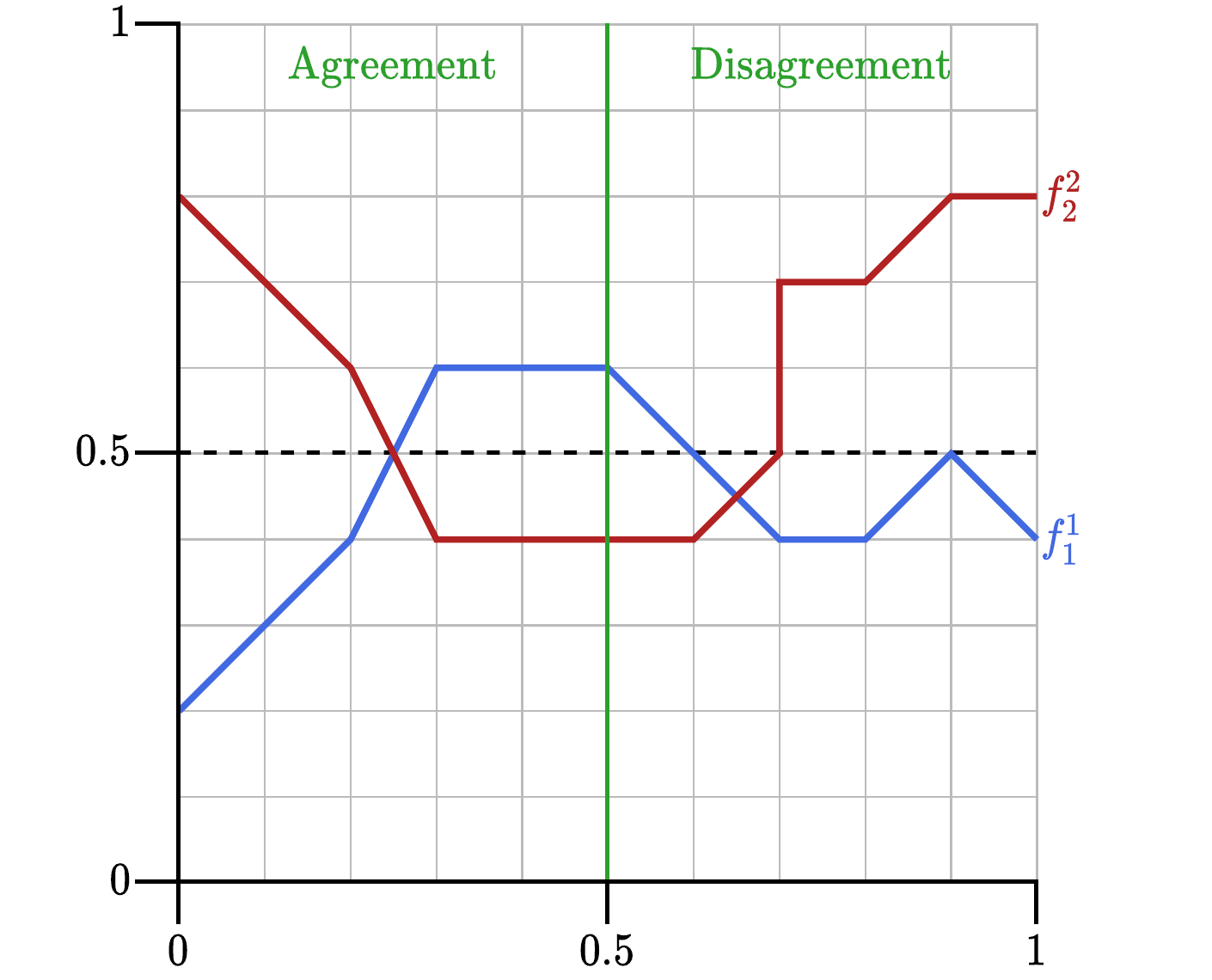}{\label{figTheoreticalExample1} An instance of the state-cutting problem where $N = \{1, 2\}$ and $m = 10$. The density functions $f_1^1$ and $f_2^2$ are shown in blue and red, respectively. This is a full specification of the instance, since we must have $f_1^2(x) = 1 - f_1^1(x)$ and $f_2^1(x) = 1 - f_2^2(x)$, and the voter distribution functions can be computed by taking integrals, e.g., $v_1^1([0.5, 0.7]) = \int_{0.5}^{0.7}f_1^1(x)dx = 0.1$. The two parties happen to agree on the distribution of voters over $[0, 0.5]$, but disagree everywhere else.}

To discuss the number of seats won by a party with respect to a partition of $[0, 1]$ into districts, we are confronted with the technical issue of how to resolve perfect ties. Our solution is to assume that whoever is drawing the electoral districts has the ability to resolve ties in whatever way they wish. In other words, a district partition comes with a built-in tie-breaking rule, so to define a partition, one must not only specify where within $[0, 1]$ each district lies, but also who wins each district in the case of a tie. Our results do not depend critically on this modeling choice; it is mainly for elegance and ease of exposition. Formally, for any $m \in Z_{\geq 1}$ and $S \subseteq [0, 1]$, an \emph{$m$-partition of $S$} is a pair $(P, T)$, where $P = \{\seq{D}{m\mu(S)}\}$ is a set of districts and $T: P \to N$ is a tie-breaking rule. Furthermore, $P$ must satisfy the following axioms:
\begin{enumerate}
	\item\label{itmPartitionMeasure} For all $k$, $\mu(D_k) = \frac{1}{m}$.
	\item\label{itmPartitionDisjoint} For all $k_1, k_2$, $\mu(D_{k_1} \cap D_{k_2}) = 0$ (i.e., districts only overlap at endpoints).
	\item\label{itmPartitionUnion} $\bigcup_{k} D_k = S$.
\end{enumerate}
We write $\mathcal{P}(m)$ for the set of all $m$-partitions of $[0, 1]$. Given an instance of the state-cutting problem and an $m$-partition $(P, T)$, we denote the number of districts won (in the sense of absolute majority) by each party $j \in N$, according to party $i \in N$, by
$$u_i^j(P, T) := {\bigg|\bigg\{}D \in P \suchthat v_i^j(D) > \frac{1}{2m} \txt{ or } \left(v_i^j(D) = \frac{1}{2m} \txt { and } T(D) = j\right){\bigg\}\bigg|}.$$
For each district $D$ in the set above, we say that party \emph{$j$ wins $D$ according to $i$ under $(P, T)$}. When $j = i$, we simply say \emph{$i$ wins $D$ under $(P, T)$}. A \emph{GT partition} is an $m$-partition $(P, T)$ of $[0, 1]$ such that, for all $i \in N$, the  \emph{geometric target for party $i$} is satisfied:
$$u_i^i(P, T) \geq \left\lfloor\frac12\left(\min\limits_{\substack{(P', T')\\\in \mathcal{P}(m)}} u_i^i(P', T') + \max\limits_{\substack{(P', T')\\\in \mathcal{P}(m)}} u_i^i(P', T')\right)\right\rfloor.$$

For example, in the instance from Figure \ref{figTheoreticalExample1}, we may define a $10$-partition $(P, T)$ by taking
\begin{align*}
	P := \{&[0, 0.1], [0.1, 0.2], [0.2, 0.3], [0.3, 0.4], [0.4, 0.5],\\
	&[0.5, 0.6], [0.6, 0.7], [0.7, 0.8], [0.8, 0.9], [0.9, 1]\}.
\end{align*}
According to party 1, party 1 only wins districts $[0.3, 0.4]$, $[0.4, 0.5]$, $[0.5, 0.6]$, and, depending on $T$, $[0.2, 0.3]$. Party 2 agrees with this assessment, except that party 1 also wins $[0.6, 0.7]$ according to party 2. As shown in Section \ref{secProofs}, the geometric target for party 1 is to win at least $\lfloor\frac{0 + 8}{2}\rfloor = 4$ districts, and the geometric target for party 2 is to win at least $\lfloor\frac{3 + 10}{2}\rfloor = 6$ districts, each according to their own voter distribution functions. Thus, if we set $T([0.2, 0.3]) := 1$, the geometric target for party 1 will be satisfied; if we set $T([0.2, 0.3]) := 2$, the geometric target for party 2 will be satisfied; but there is no choice of tie-breaking rule satisfying both targets simultaneously. In other words, for this choice of $P$, there is no $T$ such that $(P, T)$ is a GT partition.

\section{Existence of GT Partitions}\label{secProofs}
It is relatively straightforward to see that GT partitions always exist in the case where $v_1^1 \equiv v_2^1$, meaning that both parties agree exactly on the distribution of party support. The following theorem is superseded by our main result (Theorem \ref{thmUnboundedExistence}), but it is nevertheless instructive as a warm-up. We argue that is possible to transform any partition into a canonical partition through a sequence of small steps, each of which changes the balance of power by at most one. This implies a sequence of transitions from a party's worst partition to their best that does not skip any intermediate outcome and, in particular, includes a GT partition. 

\begin{theorem}\label{thmAgreementExistence}
	Given any instance of the state-cutting problem in which $N = \{1, 2\}$ and $v_1^1 \equiv v_2^1$, a GT partition always exists.
\end{theorem}

\begin{proof}
	Let $(P_1, T_1)$ be a best $m$-partition of $[0, 1]$ for party 1 (which is a worst $m$-partition for party 2), and let $(P_2, T_2)$ be a worst $m$-partition of $[0, 1]$ for party 1 (which is a best $m$-partition for party 2). Without loss of generality assume each $T_i$ breaks ties in favor of party $i$. For any given $i \in \{1, 2\}$, we imagine bubble-sorting the disjoint intervals comprising the districts of $P_i$, where the sort key of an interval is the index of the district in $P_i$ to which it belongs. Each time two adjacent intervals are swapped, we repartition the corresponding subinterval to get a new partition, as in Figure \ref{figBubbleSort}. In the end, we arrive at the simplest possible partition $P^*$, in which each district is connected (like the example $P$ from Section \ref{secModel}). This creates a chain of partitions from $P_1$ to $P^*$ to $P_2$, each differing from the previous one on at most 2 districts (the ones containing the adjacent intervals that were swapped). Consistently using $T_1$ to break ties, we have a chain of $m$-partitions from $(P_1, T_1)$ to $(P_2, T_1)$.
	
	\begin{figure}
		\includegraphics[width=\columnwidth]{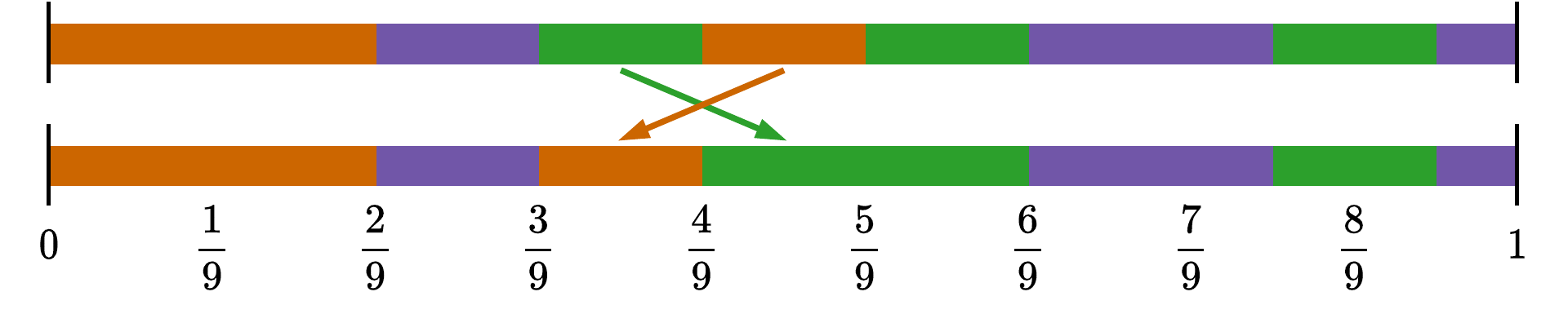}
		\caption{\label{figBubbleSort} An illustration of the repartitioning step in the proof of Thm. \ref{thmAgreementExistence} with three districts, where intervals of the same color form a district. The top partition is\\$\left\{{\color{niceColorBrown}\left[0, \frac{2}{9}\right] \cup \left[\frac{4}{9}, \frac{5}{9}\right]}, {\color{niceColorGreen}\left[\frac{3}{9}, \frac{4}{9}\right] \cup \left[\frac{5}{9}, \frac{6}{9}\right] \cup \left[\frac{15}{18}, \frac{17}{18}\right]}, {\color{niceColorPurple}\left[\frac{2}{9}, \frac{3}{9}\right] \cup \left[\frac{6}{9}, \frac{15}{18}\right] \cup \left[\frac{17}{18}, 1\right]}\right\}$, and the bottom partition is\\ $\left\{{\color{niceColorBrown}\left[0, \frac{2}{9}\right] \cup \left[\frac{3}{9}, \frac{4}{9}\right]}, \hspace{3.64em}{\color{niceColorGreen}\left[\frac{4}{9}, \frac{6}{9}\right] \cup \left[\frac{15}{18}, \frac{17}{18}\right]}, {\color{niceColorPurple}\left[\frac{2}{9}, \frac{3}{9}\right] \cup \left[\frac{6}{9}, \frac{15}{18}\right] \cup \left[\frac{17}{18}, 1\right]}\right\}$.}
	\end{figure}
	
	We claim that, at each step in this chain, the number of districts won by party 1 (and thus party 2 as well) changes by at most $\pm1$. Suppose toward a contradiction that this was not the case at some step, going from $(P, T_1)$ to $(P', T_1)$. Let $v^1$ denote the common function $v_1^1 \equiv v_2^1$. Let the two districts on which $P$ and $P'$ differ be $D_1, D_2 \in P$ and $D_1', D_2' \in P'$. Since we are breaking ties in favor of party 1, the only way that the number of wins can differ by at least 2 is if party 1 has a weak majority in $D_1$ and $D_2$, but a strict minority in $D_1'$ and $D_2'$; or a strict minority in $D_1$ and $D_2$, and a weak majority in $D_1'$ and $D_2'$. These two cases are completely analogous, so we only consider the former case, i.e., $v^1(D_1) \geq \frac{1}{2m}$, $v^1(D_2) \geq \frac{1}{2m}$, $v^1(D_1') < \frac{1}{2m}$, and $v^1(D_2') < \frac{1}{2m}$. Then, by the additivity of $v^1$,
    \begin{align*}
	\frac1m &\leq v^1(D_1) + v^1(D_2) = v^1(D_1 \cup D_2)
    = v^1(D_1' \cup D_2')\\
    &= v^1(D_1') + v^1(D_2') < \frac1m.
 \end{align*}
	We have a contradiction, so the number of districts won by party 1 can change by at most $\pm1$ at each link in the chain. 
	
	Finally, we extend the chain by $m$ more steps from $(P_2, T_1)$ to $(P_2, T_2)$ by changing the tie-breaking rule one district at a time. Again, the number of wins for party 1 changes by at most $\pm1$ at each step. Thus, at some point in the middle of the chain of $m$-partitions from $(P_1, T_1)$ to $(P_2, T_2)$, the rounded average number of wins for each party between these two extremes is realized.
\end{proof}

Our main result concerns the general case where the parties may disagree on the distribution of voters. 

\begin{theorem}\label{thmUnboundedExistence}
	Given any instance of the state-cutting problem in which $N = \{1, 2\}$, a GT partition always exists.
\end{theorem}

The proof is via an interactive protocol. It is important to note that we do not suggest using this protocol; rather, it is merely a tool for proving the feasibility of GT partitions, and this notion is then used as a constraint for optimization, as we discuss in Section~\ref{secPractice}.

The full proof is technical and broken up into several lemmas; let us first give a high-level overview. We begin by identifying what each party $j$ considers to be the ``battleground'' areas, where both parties have the same level of support, so either party could hope to win districts by gerrymandering. Formally, this is defined as a set $X_j \subseteq [0, 1]$ of maximal size such that both parties have the same level of support over $X_j$ according to party $j$'s beliefs on the distribution of voters. Our first key observation (Lemmas \ref{lemFCFMinorityValue} and \ref{lemFCFMajorityValue}) is that a best partition for party $j$ is one that perfectly divides $X_j$ into districts that $j$ barely wins, while a worst partition is one that perfect divides $X_j$ into districts that $j$ barely loses, and whatever happens outside of $X_j$ in these extreme cases is irrelevant. It follows that, in order for player $j$ to get halfway from their worst possible utility to their best possible utility, thereby satisfying their geometric target, it suffices for player $j$ to be granted control over redistricting (a specific) half of their $X_j$ set.

Thus, consider the following cut-and-choose protocol, where the party $j$ with the smaller $X_j$ set is the cutter, and the other party $i$ is the chooser. The cutter divides $X_j$ into two pieces such that they can meet their geometric target as long as they control the redistricting over \emph{either} piece. The chooser must then cede control over one of these pieces, redistricting the rest of the interval in any way they wish.

The difficult part of the proof lies in showing that the chooser $i$ will be satisfied with at least one of these two choices. To decide which piece is better, there are two different cases, depending on whether $i$ believes they are a minority or a majority party. If $i$ is a minority party, they use the partition of $X_j$ to induce a partition of $X_i$ into two equal pieces, and cede control over the piece in which they have less support, retaining control over the piece in which they have more support. It is then not too difficult to show that $i$ will be able to meet its geometric target just from forming districts within the retained half of $X_i$.

The case where $i$ is a majority party is more involved, since it may happen that, no matter which piece $D \subseteq X_j$ they cede to party $j$, it might be impossible for $i$ to form enough districts from the remains to meet their geometric target. This is because both choices let party $j$ form ``packed'' districts, in which party $i$ wins by a large margin, wasting its advantage. However, when this happens, party $i$ can respond by forming a packed district in $[0, 1] \setminus D$ that party $j$ wins for each packed district in $D$ that party $i$ wins. We argue that, for at least one of the two choices of $D$, party $i$ will be left with majority over the remainder of the interval after forming these packed districts, so will be able to win all remaining districts. Since the wins in packed districts for each party exactly cancel each other out, this implies that party $i$ meets their geometric target.

Formally, we begin by observing that it is possible to subdivide any district into two smaller districts of arbitrary sizes with the same fraction of party support as the original district. This is similar to the well-known ``Austin Cut Procedure'' from cake-cutting \cite{AustinCut}.

\begin{lemma}\label{lemCut}
	Given a voter distribution function $v$, a district $D$, and a real number $s \in [0, 1]$, there exist districts $D_1$ and $D_2$ such that
	\begin{enumerate}
		\item\label{itmCutUnion} $D_1 \cup D_2 = D$,
		\item\label{itmCutIntersection} $\mu(D_1 \cap D_2) = 0$,
		\item\label{itmCutMeasure} $\mu(D_1) = s\mu(D)$, $\mu(D_2) = (1 - s)\mu(D)$, and
		\item\label{itmCutProportional} $v(D_1) = s v(D)$, $v(D_2) = (1 - s)v(D)$.
	\end{enumerate}
\end{lemma}

By iteratively applying Lemma \ref{lemCut}, we obtain a more general form. The proof is completely straightforward, and hence omitted.

\begin{lemma}\label{lemCutIterated}
	Given a voter distribution function $v$, a district $D$, and $s \in \rr_{> 0} \cup \{\infty\}$, there exist districts $\seq{D}{\lfloor 1/s \rfloor}$ such that
	\begin{enumerate}
		\item\label{itmCutIteratedUnion} for all $k$, $D_k \subseteq D$,
		\item\label{itmCutIteratedIntersection} for all $k_1 \neq k_2$, $\mu(D_{k_1} \cap D_{k_2}) = 0$,
		\item\label{itmCutIteratedMeasure} for all $k$, $\mu(D_k) = s\mu(D)$, and
		\item\label{itmCutIteratedProportional} for all $k$, $v(D_k) = s v(D)$.
	\end{enumerate}
\end{lemma}

Throughout the remainder of this section, fix an instance of the state-cutting problem with $N = \{1, 2\}$. For any $i, j \in N$, we say that $j$ is a \emph{minority party according to $i$} if $v_i^j([0, 1]) \leq \frac12$, and a \emph{majority party according to $i$} if $v_i^j([0, 1]) \geq \frac12$. When $j = i$, we simply say $i$ is a minority/majority party. Note that this definition is merely with respect to the data of party $i$, so even if the inequalities are strict, it is still possible for both parties to be minority parties or both parties to be majority parties. Say that a district $D$ is \emph{competitive for $i$} if $v_i^j(D) = \frac{\mu(D)}{2}$ for some $j \in N$ (in which case it will clearly be true for all $j \in N$, since there are only two parties), and let
$$M_i := \{m\mu(D) \suchthat D \txt{ is a competitive district for $i$ and } m\mu(D) \in \zz\}.$$
Since $M_i$ is a nonempty set of integers that is bounded above (by $m$), it contains a maximum value. Let $m_i \in \zz_{\geq 0}$ be this maximum, and let $X_i$ be one of the districts $D$ attaining it, i.e., $m\mu(X_i) = m_i$. Note that $m_i$ might be 0, in which case $X_i$ is empty. Figure \ref{figTheoreticalExample2} shows the sets $X_1$ and $X_2$ for our running example (in this case they are both uniquely defined, up to adding sets of measure zero). Since $m = 10$, we have $m_1 = m\mu(X_1) = 7$ and $m_2 = m\mu(X_2) = 8$.

The next five lemmas characterize the best and worst partitions for each party in terms of the $m_i$ values, giving necessary and sufficient conditions for satisfying the geometric targets.

\begin{lemma}\label{lemIVT}
	For any $i, j \in N$, let $Y$ be a district such that one of
	$$v_i^j(Y) - \frac{\mu(Y)}{2} \ \ \ \ \txt{and} \ \ \ \ v_i^j([0, 1]) - \frac12$$
	is $\geq 0$ and the other is $\leq 0$. Then $\mu(Y) < \frac{m_i + 1}{m}$.
\end{lemma}

\begin{proof}
	Suppose toward a contradiction that
	$$\mu(Y) \geq \frac{m_i + 1}{m}.$$
	Define a function $g: [0, 1] \to [-\frac12, \frac12]$ by
	$$g(t) := v_i^j(Y \cup [0, t]) - \frac{\mu(Y \cup [0, t])}{2}.$$
	Clearly, $g$ is continuous. Furthermore,
	\begin{align*}
	g(0) &= v_i^j(Y) - \frac{\mu(Y)}{2},\\
	g(1) &= v_i^j([0, 1]) - \frac{\mu([0, 1])}{2} = v_i^j([0, 1]) - \frac12.
	\end{align*}
	By assumption, one of these terms must be $\geq 0$ and the other $\leq 0$. Therefore, by the intermediate value theorem, there exists $t^* \in [0, 1]$ such that $g(t^*) = 0$. Letting $D := Y \cup [0, t^*]$, we must have that
	$$v_i^j(D) = g(t^*) + \frac{\mu(D)}{2} = \frac{\mu(D)}{2},$$
	i.e., $D$ is competitive for $i$. Since $Y \subseteq D$,
	$$\mu(D) \geq \mu(Y) \geq \frac{m_i + 1}{m}.$$
	Thus, we may apply Lemma \ref{lemCut} to voter distribution function $v_i^j$, with $s := \frac{m_i + 1}{m\mu(D)} \in [0, 1]$, to cut out a district $D_1 \subseteq D$ of measure
	$$\mu(D_1) = \frac{m_i + 1}{m\mu(D)} \cdot \mu(D) = \frac{m_i + 1}{m}.$$
	Furthermore, observe that, since $D$ is competitive for $i$, it follows from property (\ref{itmCutProportional}) of Lemma \ref{lemCut} that $D_1$ is competitive for $i$:
	\begin{align*}
	v_i^j(D_1) &= s \cdot v_i^j(D)\\
	&= s \cdot \frac{\mu(D)}{2} \stext{because $D$ is competitive for $i$}\\
	&= \frac{m_i + 1}{m\mu(D)} \cdot \frac{\mu(D)}{2}\\
	&= \frac{m_i + 1}{2m}\\
	&= \frac{\mu(D_1)}{2}.
	\end{align*}
	This proves that $m_i + 1 \in M_i$, contradicting the definition of $m_i$ as the maximum element of $M_i$.
\end{proof}

\begin{lemma}\label{lemFCFMinorityValue}
	For any $i, j \in N$, if $j$ is a minority party according to $i$, then:
	\begin{align*}
	\min_{(P', T') \in \mathcal{P}(m)} u_i^j(P', T') &= 0\\
	\max_{(P', T') \in \mathcal{P}(m)} u_i^j(P', T') &= m_i
	\end{align*}
\end{lemma}

\begin{proof}
	Let $j'$ denote the party that is not $j$. To prove the first equation, we apply Lemma \ref{lemCutIterated} to $v_i^j$, with $s := \frac{1}{m}$, to divide $[0, 1]$ into $m$ districts $\seq{D}{m}$ of equal size $\frac1m$. In each district $D_k$, from property (\ref{itmCutIteratedProportional}) of Lemma \ref{lemCutIterated} and the fact that $j$ is a minority party according to $i$,
	$$v_i^j(D_k) = \frac1m v_i^j([0, 1]) \leq \frac{1}{2m}$$
	Therefore, if we break ties in favor of party $j'$, party $j$ will win none of these districts. Formally, letting $P' := \{\seq{D}{m}\}$ and $T'(D_k) := j'$ for each $k \in [m]$, we have that $u_i^j(P', T') = 0$, proving the first equation.
	
	To prove the second equation, we apply Lemma \ref{lemCutIterated} to $v_i^j$, with $s := \frac{1}{m_i}$, to divide $X_i$ into $m_i$ districts $\seq{D}{m_i}$. Note that, by property (\ref{itmCutIteratedMeasure}) of Lemma \ref{lemCut}, each district $D_k$ has size
	$$\mu(D_k) = s \cdot \mu(X_i) = \frac{1}{m_i} \cdot \frac{m_i}{m} = \frac1m.$$
	Furthermore, since $X_i$ is competitive for $i$, it follows from property (\ref{itmCutIteratedProportional}) of Lemma \ref{lemCutIterated} that each $D_k$ is competitive for $i$. Let $P'$ consist of $\seq{D}{m_i}$, along with an arbitrary division of $\overline{[0, 1] \setminus X_i}$ (the closure of the complement of $X_i$) into $m - m_i$ districts, and let $T'(D_k) := j$ for each $k \in [m_i]$, with an arbitrary tie-breaking choice for all of the other districts. Since the $D_k$ districts are competitive and ties are broken in favor of party $j$, it follows that party $j$ will win each of them according to $i$. Therefore,
	$$u_i^j(P', T') \geq m_i,$$
	which proves that
	$$\max_{(P', T') \in \mathcal{P}(m)} u_i^j(P', T') \geq m_i.$$
	
	To prove the other direction, suppose toward a contradiction that, for some $m$-partition $(P', T')$ of $[0, 1]$, $u_i^j(P', T') \geq m_i + 1$. Let $Y \subseteq [0, 1]$ be the union of all districts won by $j$ according to $i$ under $(P', T')$. Since there are at least $m_i + 1$ such districts, each of measure $\frac1m$, we have
	\begin{equation}\label{equIVTInvocationContradiction1}
	\mu(Y) \geq \frac{m_i + 1}{m}.
	\end{equation}
	However,
	$$v_i^j(Y) - \frac{\mu(Y)}{2} \geq 0$$
	since party $j$ wins each of the districts comprising $Y$ according to $i$, and
	$$v_i^j([0, 1]) - \frac12 \leq 0$$
	since party $j$ is a minority party according to $i$. Therefore, by Lemma \ref{lemIVT}, we have
	$$\mu(Y) < \frac{m_i + 1}{m},$$
	contradicting inequality (\ref{equIVTInvocationContradiction1}).
\end{proof}

\begin{lemma}\label{lemFCFMajorityValue}
	For any $i, j \in N$, if $j$ is a majority party according to $i$, then:
	\begin{align*}
	\min_{(P', T') \in \mathcal{P}(m)} u_i^j(P', T') &= m - m_i\\
	\max_{(P', T') \in \mathcal{P}(m)} u_i^j(P', T') &= m
	\end{align*}
\end{lemma}

\begin{proof}
	Let $j'$ denote the party that is not $j$. Note that $j'$ must be a minority party according to $i$. For any $m$-partition $(P', T')$ of $[0, 1]$,
	$$u_i^j(P', T') + u_i^{j'}(P', T') = m.$$
	Therefore,
	\begin{align*}
	\min_{(P', T') \in \mathcal{P}(m)} u_i^j(P', T') &= \min_{(P', T') \in \mathcal{P}(m)} \left(m - u_i^{j'}(P', T')\right)\\
	&= m - \max_{(P', T') \in \mathcal{P}(m)} u_i^{j'}(P', T')\\
	&= m - m_i,
	\end{align*}
	where the final equality follows from Lemma \ref{lemFCFMinorityValue} and the fact that $j'$ is a minority party according to $i$. By the same reasoning, we analogously derive
	\begin{align*}
	\max_{(P', T') \in \mathcal{P}(m)} u_i^j(P', T') &= \max_{(P', T') \in \mathcal{P}(m)} \left(m - u_i^{j'}(P', T')\right)\\
	&= m - \min_{(P', T') \in \mathcal{P}(m)} u_i^{j'}(P', T')\\
	&= m. \qedhere
	\end{align*}
\end{proof}

\begin{lemma}\label{lemFCFMinoritySufficientCondition}
	For any $i \in N$ and $m$-partition $(P, T)$ of $[0, 1]$, if party $i$ is a minority party, then $(P, T)$ satisfies the geometric target for $i$ if and only if $i$ wins at least $\left\lfloor\frac{m_i}{2}\right\rfloor$ districts under $(P, T)$.
\end{lemma}

\begin{proof}
	This follows immediately from specializing $j := i$ in Lemma \ref{lemFCFMinorityValue}, since the geometric target is for party $i$ to win at least
	$$\left\lfloor\frac{0 + m_i}{2}\right\rfloor = \left\lfloor\frac{m_i}{2}\right\rfloor$$
	districts.
\end{proof}

\begin{lemma}\label{lemFCFMajoritySufficientCondition}
	For any $i \in N$ and $m$-partition $(P, T)$ of $[0, 1]$, if party $i$ is a majority party, then $(P, T)$ satisfies the geometric target for $i$ if and only if $i$ wins at least $m - \left\lceil\frac{m_i}{2}\right\rceil$ districts under $(P, T)$.
\end{lemma}

\begin{proof}
	This follows from specializing $j := i$ in Lemma \ref{lemFCFMajorityValue}, since the geometric target is for party $i$ to win at least
	$$\left\lfloor\frac{(m - m_i) + m}{2}\right\rfloor = \left\lfloor\frac{2m - m_i}{2}\right\rfloor = \left\lfloor m - \frac{m_i}{2}\right\rfloor = m + \left\lfloor\frac{-m_i}{2}\right\rfloor = m - \left\lceil\frac{m_i}{2}\right\rceil$$
	districts.
\end{proof}

\begin{lemma}\label{lemFCFCompetitiveSufficientCondition}
	For any $i \in N$ and $m$-partition $(P, T)$ of $[0, 1]$, if party $i$ wins at least $\left\lfloor\frac{m_i}{2}\right\rfloor$ competitive districts under $(P, T)$, then $(P, T)$ satisfies the geometric target for $i$.
\end{lemma}

\begin{proof}
	Let $j$ denote the party that is not $i$. If $i$ is a minority party, the result follows immediately from Lemma \ref{lemFCFMinoritySufficientCondition}. If $i$ is a majority party, then, by Lemma \ref{lemFCFMajoritySufficientCondition}, the geometric target is for party $i$ to win at least $m - \left\lceil\frac{m_i}{2}\right\rceil$ districts. Suppose toward a contradiction that $(P, T)$ did not meet the geometric target for $i$, i.e., $i$ wins strictly less than $m - \left\lceil\frac{m_i}{2}\right\rceil$ districts under $(P, T)$. Let $(P', T')$ be the $m$-partition of $[0, 1]$ where $P' := P$ and $T'(D) := j$ for all $D \in P'$. With the new tie-breaking rule $T'$, each of the $\left\lfloor\frac{m_i}{2}\right\rfloor$ competitive districts that party $i$ won under $(P, T)$ are instead won by party $j$ according to $i$ under $(P', T')$. Thus, party $i$ wins $\left\lfloor\frac{m_i}{2}\right\rfloor$ fewer districts under $(P', T')$, which is strictly less than
	$$\left(m - \left\lceil\frac{m_i}{2}\right\rceil\right) - \left\lfloor\frac{m_i}{2}\right\rfloor = m - m_i$$
	districts in total. This contradicts the minimum value from Lemma \ref{lemFCFMajorityValue}.
\end{proof}

We are now ready to prove the main result. Below, we only consider the simpler case where the chooser party $i$ is a minority party; the majority case is deferred to Appendix \ref{appProofMajorityCase}.

\begin{proof}[Proof of Theorem \ref{thmUnboundedExistence}]
	Choose $i, j \in N = \{1, 2\}$ so that $i \neq j$ and $m_i \geq m_j$. (Party $j$ will be the cutter, and party $i$ will be the chooser.) We first apply Lemma \ref{lemCut} to voter distribution function $v_j^j$, on district $X_j$, with $s := \frac12$, obtaining districts $D_1$ and $D_2$ satisfying the four properties. See Figure \ref{figTheoreticalExample2} for an example of one valid choice of $D_1$ and $D_2$. Note that, for each $k \in \{1, 2\}$, from property (\ref{itmCutMeasure}) of Lemma \ref{lemCut} we have
	\begin{equation}\label{equD1D2Measure}
	\mu(D_k) = \frac{\mu(X_j)}{2} = \frac{m_j}{2m},
	\end{equation}
	while from property (\ref{itmCutProportional}), $D_k$ is competitive for $j$ since $X_j$ is.
	
	\ipncm{0.65}{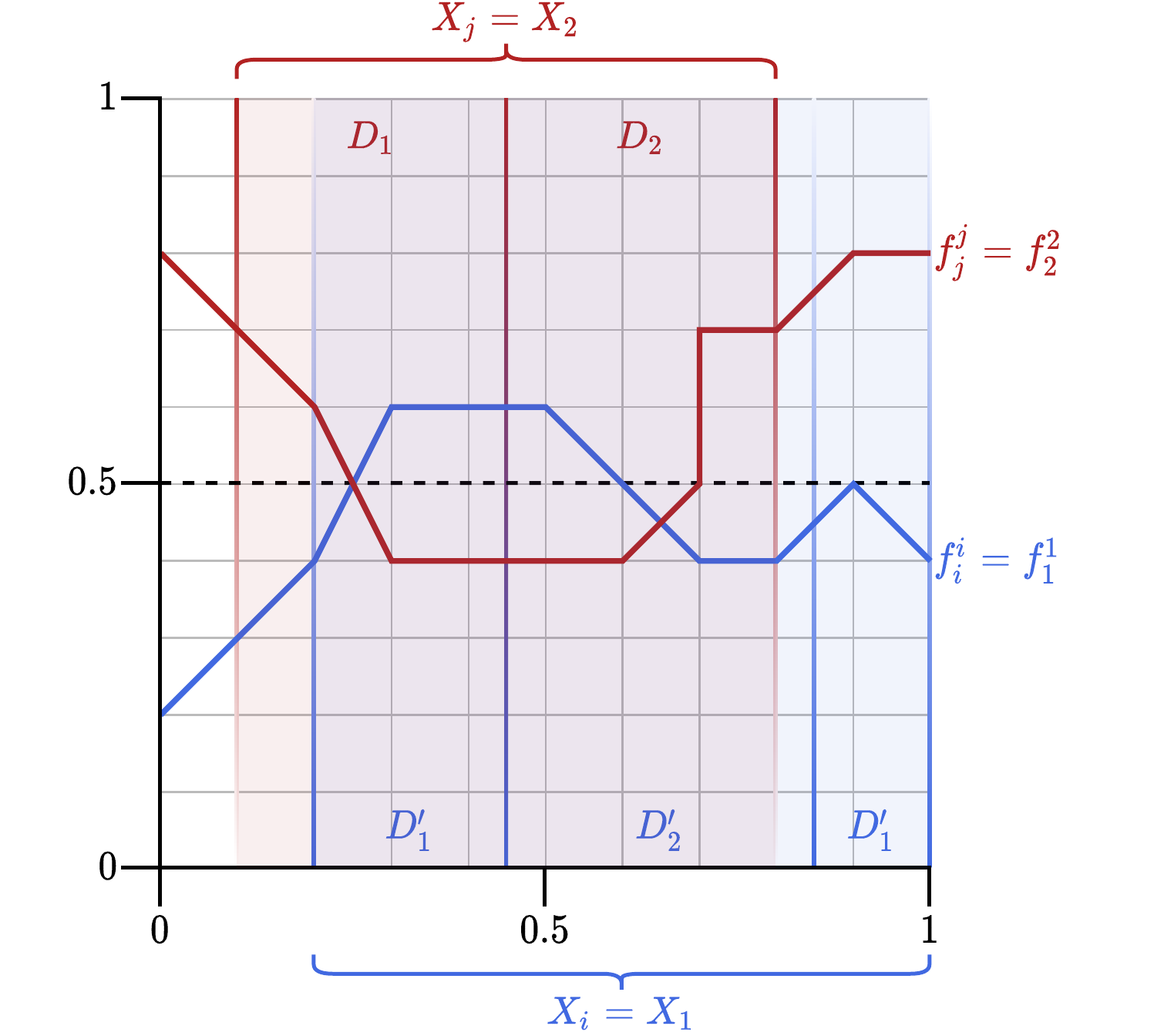}{\label{figTheoreticalExample2} The same instance of the state-cutting problem from Figure \ref{figTheoreticalExample1}, annotated with some of the sets described in the proof of Theorem \ref{thmUnboundedExistence}. Note that $i = 1$ and $j = 2$ since $\mu(X_1) > \mu(X_2)$.}
	
	We claim that, for any $k_j \in \{1, 2\}$, it is possible to create an $m$-partition of a subset of $D_{k_j}$ such that, no matter how this partition is extended into an $m$-partition of $[0, 1]$, the geometric target for party $j$ is satisfied.
	
	To prove this, we apply Lemma \ref{lemCutIterated}, to $v_j^j$, with $s := \frac{2}{m_j}$, to cut $\left\lfloor\frac{m_j}{2}\right\rfloor$ districts
	$$P_{k_j} := \left\{E_{1}, E_{2}, \dots, E_{\left\lfloor\frac{m_j}{2}\right\rfloor}\right\}$$
	from $D_{k_j}$. From property (\ref{itmCutIteratedMeasure}) of Lemma \ref{lemCutIterated}, for each $E_{k}$ district,
	\begin{align*}
	\mu(E_{k}) &= s\mu(D_{k_j})\\
	&= \frac{2}{m_j} \cdot \frac{m_j}{2m} \stext{from equation (\ref{equD1D2Measure})}\\
	&= \frac1m,
	\end{align*}
	and, from property (\ref{itmCutIteratedProportional}), each of these districts is competitive for $j$ since $D_{k_j}$ was. Thus, defining the tie-breaker over each $E_{k}$ district by $T_{k_j}(E_{k}) := j$ ensures that party $j$ wins all of these $\left\lfloor\frac{m_j}{2}\right\rfloor$ competitive districts under $(P_{k_j}, T_{k_j})$, so any extension of $(P_{k_j}, T_{k_j})$ satisfies the geometric target for $j$ by Lemma \ref{lemFCFCompetitiveSufficientCondition}.
	
	It thus remains to establish that, for some $k_j \in \{1, 2\}$, we can extend $(P_{k_j}, T_{k_j})$ to an $m$-partition of $[0, 1]$ that satisfies the geometric target for party $i$. There are two cases, depending on whether party $i$ is a minority or majority party (according to $i$).
	
	Suppose $i$ is a minority party, as is the case in Figure \ref{figTheoreticalExample2}. From equation (\ref{equD1D2Measure}) it follows that, for all $k \in \{1, 2\}$,
	$$\mu(D_k \cap X_i) \leq \mu(D_k) = \frac{m_j}{2m} \leq \frac{m_i}{2m} = \frac{\mu(X_i)}{2}.$$
	Therefore, it is possible to enlarge $D_1 \cap X_i$ and $D_2 \cap X_i$ into districts $D_1', D_2' \subseteq X_i$ that exactly partition $X_i$ (ignoring overlapping endpoints of measure zero), both having equal measure
	\begin{equation}\label{equDkPrimeMeasure}
	\mu(D_k') = \frac{m_i}{2m}
	\end{equation}
	(see Figure \ref{figTheoreticalExample2} for an example of a valid choice of $D_1'$ and $D_2'$).
	
	Since $X_i$ is competitive for $i$,
	$$0 = v_i^i(X_i) - \frac{\mu(X_i)}{2} = v_i^i(D_1') + v_i^i(D_2') - \frac{m_i}{2m} = \left(v_i^i(D_1') - \frac{m_i}{4m}\right) + \left(v_i^i(D_2') - \frac{m_i}{4m}\right).$$
	Therefore, the two terms in parentheses cannot both be negative. Let $k_i \in \{1, 2\}$ be such that
	\begin{equation}\label{equDkPrimeCompetitiveForI}
	v_i^i(D_{k_i}') \geq \frac{m_i}{4m},
	\end{equation}
	and let $k_j \in \{1, 2\}$ be the other index, so $k_i \neq k_j$ (in Figure \ref{figTheoreticalExample2}, $k_i = 1$ and $k_j = 2$). We construct an $m$-partition $(P'_{k_i}, T'_{k_i})$ by applying Lemma \ref{lemCutIterated}, to $v_i^i$, with $s := \frac{2}{m_i}$, to cut $\left\lfloor\frac{m_i}{2}\right\rfloor$ districts
	$$P'_{k_i} := \left\{\seq{F}{\left\lfloor\frac{m_i}{2}\right\rfloor}\right\}$$
	from $D_{k_i}'$. According to property (\ref{itmCutIteratedMeasure}), each district $F_{k}$ does indeed have the target size of
	\begin{align*}
	\mu(F_{k}) &= s\mu(D_{k_i}')\\
	&= \frac{2}{m_i} \cdot \frac{m_i}{2m}\stext{from equation (\ref{equDkPrimeMeasure})}\\
	&= \frac1m.
	\end{align*}
	Furthermore, from property (\ref{itmCutIteratedProportional}), each district $F_{k}$ has party support
	\begin{align*}
	v_i^i(F_{k}) &= s \cdot v_i^i(D_{k_i}')\\
	&= \frac{2}{m_i} \cdot v_i^i(D_{k_i}')\\
	&\geq \frac{2}{m_i} \cdot \frac{m_i}{4m} \stext{from inequality (\ref{equDkPrimeCompetitiveForI})}\\
	&= \frac{1}{2m}.
	\end{align*}
	
	\ipncm{0.65}{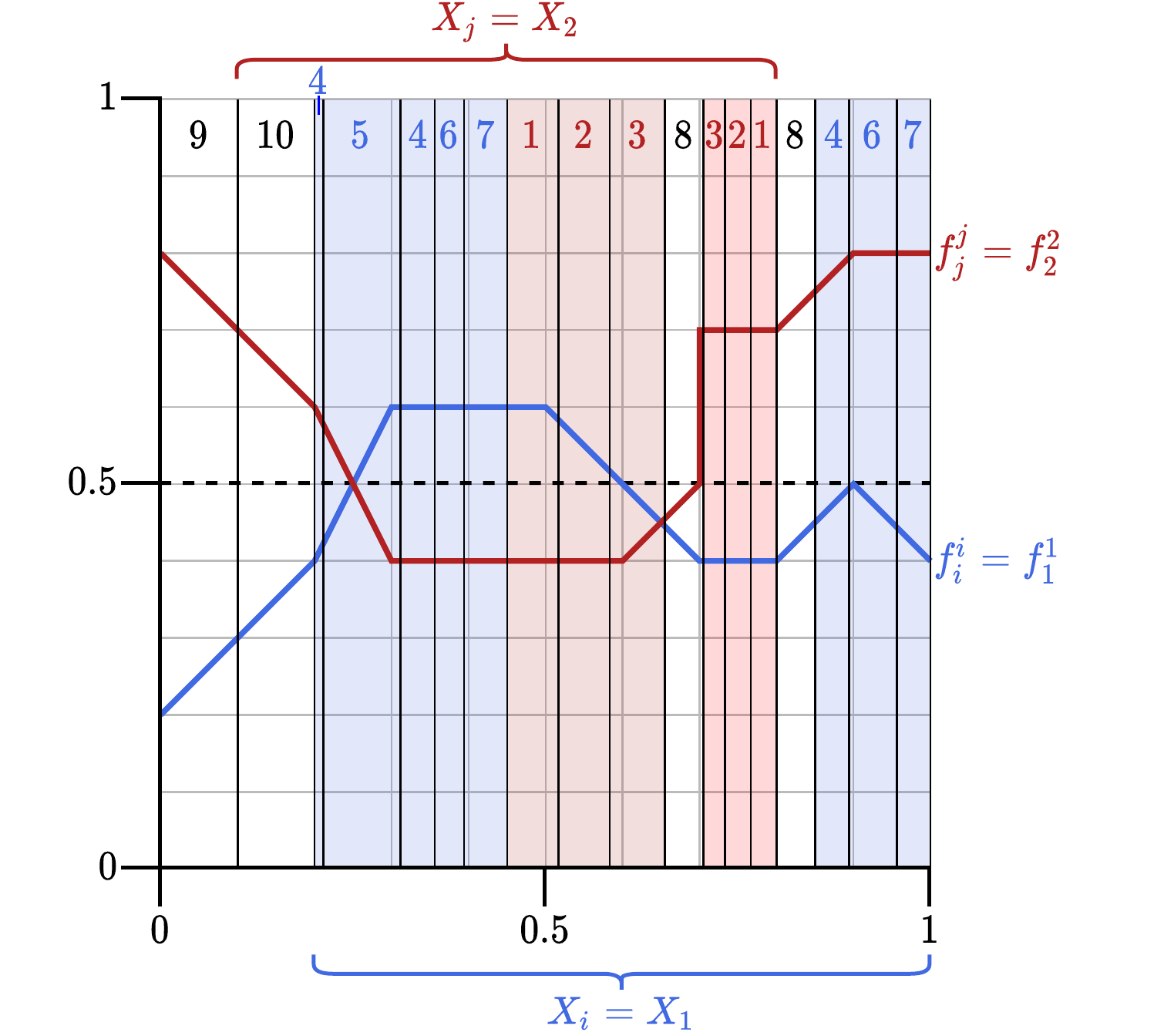}{\label{figTheoreticalExample3} The final $10$-partition meeting the geometric targets of both parties, with districts numbered in the order they are constructed in the proof of Theorem \ref{thmUnboundedExistence}. The red districts 1-3 come from $(P_{k_j}, T_{k_j})$, so have ties broken in favor of party $j = 2$, while the blue districts 4-7 come from $(P'_{k_i}, T'_{k_i})$, so have ties broken in favor of party $i = 1$ (though in this case, it does not matter, since districts 4-7 are not competitive). The white districts 8-10 could be re-partitioned arbitrarily, and have ties broken in any way. Party $i = 1$ expects to win districts 1, 4, 5, 6, and 7, exceeding their geometric target of four districts, while party $j = 2$ expects to win all except district 5, exceeding their geometric target of six districts.}
	
	We define the tie-breaker over each $F_{k}$ district by $T'_{k_i}(F_{k}) := i$, ensuring that party $i$ wins all of these $\left\lfloor\frac{m_i}{2}\right\rfloor$ districts. To form a GT partition for $[0, 1]$, we take all districts and tie-breakers from $(P'_{k_i}, T'_{k_i})$ and $(P_{k_j}, T_{k_j})$ (which are necessarily disjoint since $D_{k_i}'$ and $D_{k_j}$ are), dividing the remainder of $[0, 1]$ arbitrarily. Since party $i$ is the minority party and wins at least $\left\lfloor\frac{m_i}{2}\right\rfloor$ districts, the geometric target for party $i$ is satisfied by Lemma \ref{lemFCFMinoritySufficientCondition}. Figure \ref{figTheoreticalExample3} shows the final $10$-partition for our running example. The case where $i$ is a majority party is handled in Appendix \ref{appProofMajorityCase}.
\end{proof}

\section{GT partitions in practice}\label{secPractice}

Having established the existence of GT partitions in an abstract model, now we empirically investigate whether GT partitions exist in practice and what they look like. In the spirit of the \emph{price of fairness}~\cite{CKKK09,BFT11}, we are particularly interested in the trade-off between satisfying the geometric target and various  optimization objectives; that is, we investigate to what degree GT partitions are inferior to those that optimize traditional measures of quality.

\begin{table*}[ht]
	\small
	\centering
	\begin{tabular}{lcccccc}
		\hline 
		&	\textbf{GA}  	&	\textbf{MA}	&	 \textbf{NC} 	&	 \textbf{PA} 	&	  \textbf{TX} 	&	 \textbf{VA}   	\\ \hline
		\textbf{\# Districts} 	&	14	&	9	&	13	&	18	&	36	&	11	\\
		\textbf{Democratic vote share (\%)} 	&	47.6	&	64.7	&	48.1	&	49.6	&	45.3	&	52.5	\\
		\textbf{Democratic GT} 	&	4	&	9	&	5	&	7	&	15	&	7	\\
		\textbf{Republican GT} 	&	9	&	0	&	8	&	11	&	21	&	4	\\
		\textbf{Competitive districts} 	&	7 	&	2 	&	8 	&	8	&	12 (13)	&	6 	\\
		\textbf{Efficiency gap (\%)} 	&	6.0 (0*)	&	20.7	&	4.1 (0*)	&	5.4 (1.2)	&	0.1 (0*)	&	4.2 (0*)	\\
		\textbf{Compactness (PP)} 	&	0.214 	&	0.354 	&	0.262 	&	0.222 (0.225)	&	0.194 (0.2)	&	0.25 	\\
		\hline \vspace{.1cm}
	\end{tabular}
	\caption{\label{tab}For each state, its number of Congressional districts, the normalized Democratic vote share in the 2016 presidential election (calculated from the numbers published in the New York Times by discarding votes for third-party candidates), the Democratic and Republican geometric targets, and for each of three optimization objectives, the optimal value subject to satisfying the geometric target and the optimal value without this constraint (in parentheses, where different). Absolute efficiency gaps of $0^*$ do not exceed 0.05\%.}
\end{table*}

\igncm{.35}{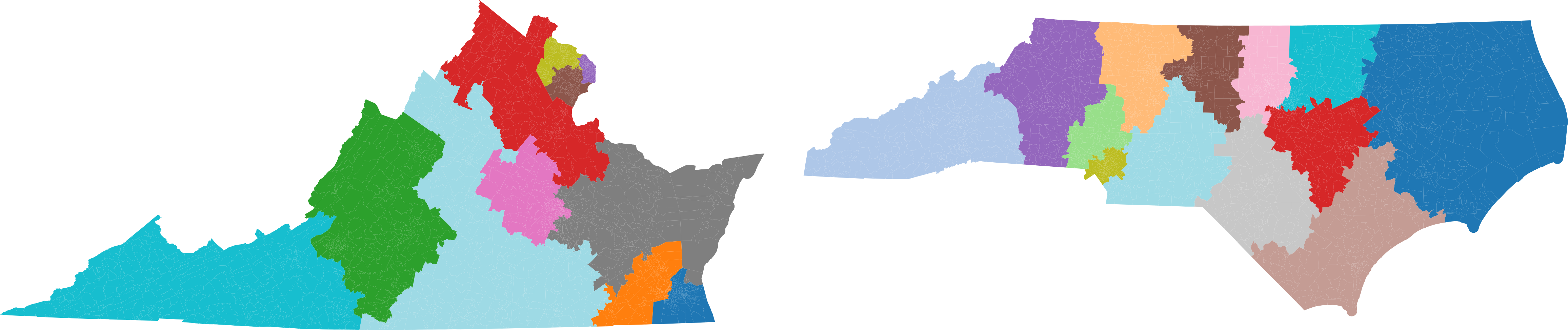}{\label{fig:mop}GT partitions in Virginia (left) and North Carolina (right) which outperform their implemented plans  in terms of competitiveness, efficiency gap and compactness.}

A first challenge, though, is computation. Ideally, we would like to exactly optimize for the number of districts each party can win and use these optimal solutions to compute the geometric targets. Unfortunately, state-of-the-art machinery does not support exact optimization over the entire space of feasible partitions at the scale of real-world instances. We therefore rely on a heuristic evaluation of the extreme partitions; specifically, we use the GerryChain software developed by the Voting Rights Data Institute \cite{gerrychain} to facilitate the running of a Markov chain which generates thousands of valid partitions. The Markov chain starts from  a graph representation of the state in which every node represents an indivisible geographic region (for example, a precinct or census block), along with properties associated with that region, including  population, area, perimeter, and the number of Democratic and Republican votes cast in several recent elections. State transitions in the Markov chain happen through \emph{recombination} moves \cite{DDS19} which merge two adjacent districts before randomly splitting them again.  Before a move to a new partition is accepted, it is verified that the new partition is contiguous and satisfies population equality to within 2\% (with the exception of Virginia, where a bound of 5\% is used).  The precinct geometries and election data used in these experiments were prepared by the MGGG Redistricting Lab and are publicly available \cite{MGGG}. 

We generate $50\,000$ valid partitions\footnote{This relatively small number of steps in the Markov chain is due to the fact that we are using recombination moves. If smaller, more local moves were used to traverse the space of partitions, several million would have been required~\cite{mggg2}.}
(of which the first $1\,000$ are discarded)
in  six  U.S.~states: Georgia (GA), Massachusetts (MA), North Carolina (NC), Pennsylvania (PA), Texas (TX), and Virginia (VA). 
At every partition found by the Markov chain we keep  track of three metrics: \begin{itemize}
	\item The efficiency gap~\cite{SM15}, which measures the net difference in the fraction of each party’s wasted votes. Every vote
cast for the minority in a district is deemed to have been wasted, as are all votes for the majority
above the threshold required to win the district. 
	\item  The number of competitive districts, defined to be those districts in which the majority party wins no more than 54\% support. 
	\item Compactness as measured by the Polsby-Popper (PP) score \cite{PP91}, computed as the ratio of the area of a district to the area of a circle with the same perimeter length.
\end{itemize}
Note that a smaller efficiency gap is better\,---\,a threshold of 8\% is commonly accepted~\cite{SM15}\,---\,while we prefer a larger number of competitive districts and a larger Polsby-Popper score.

Along with these metrics we compute the number of districts won by each party.
This allows us to calculate the geometric targets and measure the price of fairness. 

\begin{figure*}[t]
	\centering
	\includegraphics[trim=0 0 0 0.5cm, clip,scale=.41]{{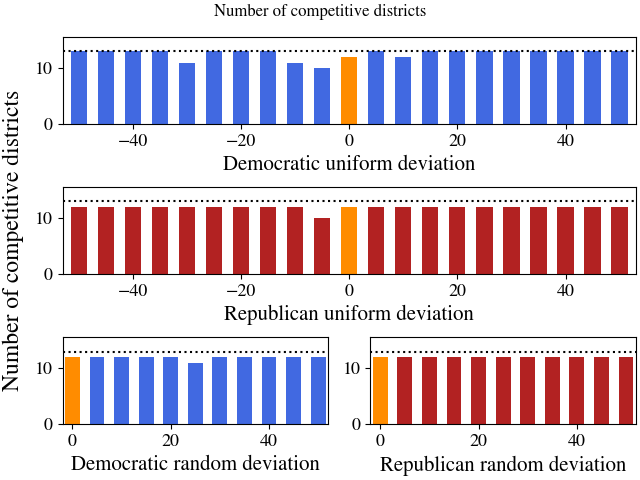}}\hspace{4mm}
	\includegraphics[trim=0 0 0 0.5cm, clip,scale=.41 ]{{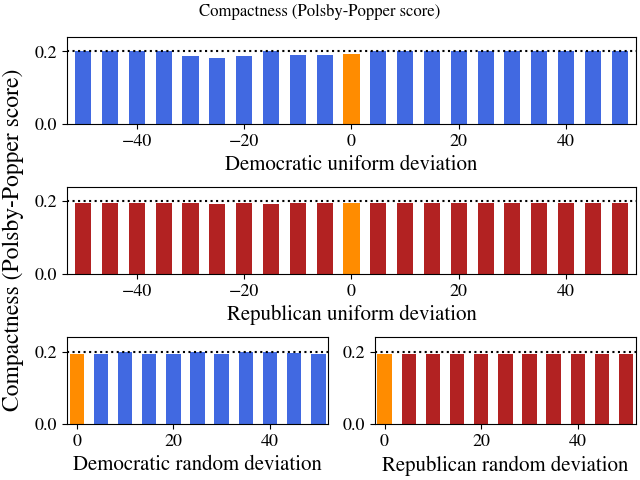}}
	\caption{The largest number of competitive districts (left panel) and degree of compactness (right) of the best GT partitions observed in Texas when parties inflate or deflate their reported voter distribution by up to 50\%. The black dotted line represents the maximum value observed among any partition. The color of the bar represents which party deviates. The golden bars report the experiment where neither party deviates, which is also recorded in the TX column of Table \ref{tab} (so in each panel, all four golden bars represent the same experiment).}
	\label{fig:dev}
\end{figure*}

\subsection{When Parties Agree About Voter Distributions}
First, we consider the case where both parties agree about the distribution of voters. In this case we use the votes cast in the 2016 presidential election to evaluate the number of districts won by each party in every partition. {In all of our experiments, we find that GT partitions exist.} Table~\ref{tab} reports the best observed values for each metric among GT partitions, as well as the optimal value observed among all partitions (when different). Geometric targets are computed by taking the average (rounded down) of the minimum and maximum number of districts won by a party in any partition of the ensemble. 

We see in Table~\ref{tab} that the cost of enforcing the geometric target is very low. There is only one instance of a state (TX) in which this constraint leads to a decrease in the number of competitive districts compared to the maximum competitive districts observed, while the decrease in compactness is never more than 3\%. The increase in efficiency gap is larger (4-6\%); however, we observe GT partitions meeting the recommended efficiency gap threshold of 8\% in every state (except MA, where meeting the threshold is impossible).

We did not explicitly consider optimizing multiple objectives simultaneously; nevertheless, we   observe several GT partitions that outperform the currently implemented partitions in these states on all three axes. Figure~\ref{fig:mop} shows two such GT partitions, one in Virginia and one in North Carolina. The Virginian partition has three competitive districts (compared to two in their 2012 plan), an absolute efficiency gap of 6.6\% (compared to 10.9\%) and a compactness score of 0.185 (compared to 0.158). Similarly, the partition of North Carolina has three competitive districts (compared to 0 in their 2016 map), an efficiency gap of 7.1\% (compared to 22.2\%) and a Polsby-Popper score of 0.262 (compared to 0.252). The implemented plans are not only worse according to all three of our metrics,   they also do not satisfy the geometric targets. 

\subsection{When Parties Disagree About Voter Distributions}\label{exp:disagree}
The core strength of our theoretical result is that it does not require parties to agree on how voters will vote, as geometric targets can be guaranteed with respect to separate beliefs for each party. {These divergent beliefs may be due to noisy data collection, polling errors or strategic manipulation.}

To simulate such settings, we consistently let one of the parties report the true votes cast in the 2016 Presidential election, which we treat as the ground truth for the purpose of computing competitiveness and efficiency gaps. The other party's beliefs are allowed to deviate in several structured ways. First, we consider the case where the other party expects the votes to reflect the 2012 Presidential election.\footnote{With the exception of Georgia and Virginia. Due to the availability of data we use election results from a 2016 senate race as the alternative voter distribution in Georgia, and the 2016 congressional races in Virginia. } 
Second, {in an attempt to simulate possible strategic behavior}, we consider what happens when the party uniformly under or over-reports their share of the votes in every region by $x$\%, for $x\in X= \{5, 10, \ldots, 50\}$. Finally, we consider the case where a party randomly inflates or deflates their share of the votes in each region (independently) by $y$\%, with $y\sim \text{Uniform}(-x, x), x \in X. $

As in the case where parties agree, in all of our experiments, we find that GT partitions exist. Figure~\ref{fig:dev} compares the most competitive and compact GT partitions observed in Texas for each of the deviations we consider. In most of the scenarios, enforcing the geometric target led to the loss of at most one competitive district; the largest number of competitive districts lost was 3. In terms of compactness (measured by the Polsby-Popper score) the largest loss was when the Democratic party deflated their reported beliefs uniformly by 25\%, leading to a GT partition with a compactness score of 0.183 compared to the optimum of 0.200. The same trends held in the setting where the alternative voter distribution is from a different election. The effect of enforcing the geometric targets on competitiveness and compactness are similar in the other states, and we observed GT partitions meeting the efficiency gap threshold everywhere (with the obvious exception of MA). The full results from all experiments appear in Appendix \ref{appEmpirical}.

Together these results tell a compelling story: not only is it easy to find GT partitions, but restricting our search to GT partitions has little impact on the quality of the partition according to traditional metrics.

\section{Discussion}\label{subDiscussion}

Our suggested redistricting approach relies on optimization subject to a fairness constraint. The fact that our fairness notion is readily satisfied 
creates the opportunity to use it in isolation should optimization-based approaches prove impossible, either because of political objections or legislative difficulties. In such cases simply requiring that partitions meet the geometric target prevents the most extreme partisan outcomes yet allows legislators to retain much of the power and freedom that comes with the ability to decide where to draw district boundaries.

Through our state-cutting model, we have demonstrated how the powerful tools of fair division can be applied to the critically important problem of political redistricting. Previous theoretical investigations of fair redistricting have been stymied by modelling issues: geometric constraints are hard to justify and intractable to work with, so typically theorems are only proved in the trivial ``geometry-free'' model where there are no constraints whatsoever. We believe our state-cutting model strikes a useful balance between these extremes, distilling the key challenges of redistricting without explicitly considering geometry. It is a fertile ground on which fairness principles for redistricting can be rigorously tested. The intuitive geometric target criterion is one such principle, though we envision more to follow. 

Nevertheless, incorporating more geometry into the model could be an interesting direction for future work. One natural way we could hope to capture ``compactness'' constraints is to count the number of disjoint intervals per district. Unfortunately, in the worst case, GT partitions may require an arbitrarily large number of intervals in some districts. Furthermore, imposing geometric constraints on the set of feasible partitions, such as ``there must be at most two intervals per district'' can lead to situations where GT partitions do not exist.\footnote{If the two parties agree on voter distributions, such counterexamples cannot occur, since the proof of Theorem \ref{thmAgreementExistence} constructs an $m$-partition with no more intervals per district than the two extreme $m$-partitions.} Perhaps there is a 2D adaptation of our model under which geometric constraints are still compatible with the geometric target.

A shortcoming of our approach is the issue of computation. A specific problem is that using the minimum and maximum number of seats won by both parties across sampled partitions to compute the geometric targets does not necessarily lead to the true value: in theory, there could be more extreme partitions that were not observed. However, this seems highly unlikely in practice. Regardless, we envision a process by which each party submits what it believes to be its best partition; the partitions submitted by the two parties can then be used to compute the geometric target of each party. Under such a process, neither party would have a right to complain that it was disadvantaged in the computation of the geometric target.  

Computation of GT partitions can also be incorporated into our theoretical model. We suspect that Robertson-Webb evaluation/cut queries~\cite{RW98} are insufficient to compute GT partitions, since it seems impossible to even compute the best and worst $m$-partitions for each party using this information, and thus it may be impossible to compute the $X_i$ sets, which form the starting point of our protocol. Is there a richer query model under which it is possible to compute a GT partition using a finite number of queries?

Another limitation of our work is that it only applies to the case of two parties. The first obstacle to extending beyond two parties is conceptual: it is unclear what the analogue of the geometric target is in that setting. We do not view this as a major issue, though, as our work is directly motivated by the process of redistricting in the United States, which essentially has a two-party system.

These shortcomings notwithstanding, our results show that it is possible and practical to guarantee fairness even in a climate of extreme partisanship. This is an insight that, we believe, could prove useful not just to academics, but also to state legislatures, courts, and independent redistricting commissions.
 
\bibliographystyle{ACM-Reference-Format}
\bibliography{mybib}

\newpage

\appendix

\section{Proof of Lemma \ref{lemCut}}\label{appProofLemCut}

    Let $f$ denote the density function of $v$. Without loss of generality, we may assume that $D = [0, t]$ where $t = \mu(D)$, for otherwise we could simply rearrange the finite number of intervals comprising $D$ so that this is the case and adapt the proof accordingly. Define functions $g: [0, 2t] \to [0, 1]$ by
	$$g(x) := \twocases{\txt{if } x \leq t}{f(x)}{\txt{if } x > t}{f(x - t)}$$
	and $h: [0, t] \to [0, 1]$ by
	$$h(x) := \int_x^{x + st} g(y) dy.$$
	Intuitively, for any $x \in [0, t]$, $h(x)$ is the value of a piece of measure $st$ that begins at $x$, wrapping around if necessary. Observe that the average value of $h$ over $[0, t]$ is
	\begin{align*}
	\frac{1}{t} \int_0^t h(x) dx &= \frac{1}{t} \int_0^t \int_x^{x + st} g(y) dy dx\\
	&= \frac{1}{t} \int_0^t \int_0^{st} g(x + y) dy dx\\
	&= \frac{1}{t} \int_0^{st} \left( \int_0^t g(x + y) dx \right) dy\\
	&= \frac{1}{t} \int_0^{st} \left( \int_y^{t + y} g(x) dx \right) dy\\
	&= \frac{1}{t} \int_0^{st} \left( \int_y^t g(x) dx + \int_t^{t + y} g(x) dx \right) dy\\
	&= \frac{1}{t} \int_0^{st} \left( \int_y^t g(x) dx + \int_0^{y} g(x + t) dx \right) dy\\
	&= \frac{1}{t} \int_0^{st} \left( \int_y^t f(x) dx + \int_0^{y} f(x) dx \right) dy\\
	&= \frac{1}{t} \int_0^{st} \left( \int_0^t f(x) dx \right) dy\\
	&= \frac{1}{t}(st) \int_0^t f(x) dx\\
	&= s v(D).
	\end{align*}
	Since $h$ is clearly continuous, by the intermediate value theorem there must exist some $x^* \in [0, t]$ at which $h$ attains its average value. If $x^* + st \leq t$, then we define
	$$D_1 := [x^*, x^* + st].$$
	In this case,
	$$\mu(D_1) = st = s\mu(D)$$
	and
	$$v(D_1) = \int_{x^*}^{x^* + st} f(y) dy = \int_{x^*}^{x^* + st} g(y) dy = h(x^*) = sv(D).$$
	If $x^* + st > t$, we instead define
	$$D_1 := [x^*, t] \cup [0, x^* + st - t].$$
	Note that these intervals are both contained within $D = [0, t]$ and are non-overlapping since $s \leq 1$ (except possibly at the point $x^*$ in the case where $s = 1$). Therefore,
	$$\mu(D_1) = (t - x^*) + (x^* + st - t) = st = s\mu(D)$$
	and
	\begin{align*}
	v(D_1) &= \int_{x^*}^{t} f(y) dy + \int_{0}^{x^* + st - t} f(y) dy\\
	&= \int_{x^*}^{t} f(y) dy + \int_{t}^{x^* + st} f(y - t) dy\\
	&= \int_{x^*}^{t} g(y) dy + \int_{t}^{x^* + st} g(y) dy\\
	&= \int_{x^*}^{x^* + st} g(y) dy\\
	&= h(x^*)\\
	&= sv(D).
	\end{align*}
	
	Thus, in either case, we have found a district $D_1$ satisfying properties (\ref{itmCutMeasure}) and (\ref{itmCutProportional}). Letting
	$$D_2 := \overline{D \setminus D_1}$$
	(the closure of $D \setminus D_1$), properties (\ref{itmCutUnion}) and (\ref{itmCutIntersection}) are automatically satisfied. Furthermore,
	$$\mu(D_2) = \mu(D) - \mu(D_1) = \mu(D) - s\mu(D) = (1 - s)\mu(D)$$
	and
	$$v(D_2) = v(D) - v(D_1) = v(D) - sv(D) = (1 - s)v(D),$$
	so $D_2$ satisfies properties (\ref{itmCutMeasure}) and (\ref{itmCutProportional}) as well. \qed

\section{Proof of Theorem \ref{thmUnboundedExistence}, majority case}\label{appProofMajorityCase}
	
	In the case where $i$ is a majority party, we first extend $(P_1, T_1)$ and $(P_2, T_2)$ by adding disjoint districts of size $\frac1m$ to $(P_1, T_1)$ and $(P_2, T_2)$, in alternation, until the total measure covered by $P_1 \cup P_2$ is exactly $\mu(X_i) = \frac{m_i}{m}$ (this is possible since $m_j \leq m_i$). Call the resulting partitions $(P'_1, T'_1)$ and $(P'_2, T'_2)$. For each $k \in \{1, 2\}$, let $A_k, B_k \subseteq [0, 1]$ be comprised of all districts that party $i$ wins/loses under $(P'_k, T'_k)$, respectively. Note that $A_1$, $A_2$, $B_1$, and $B_2$ are pairwise disjoint, have measures that are integer multiples of $\frac1m$, and for each $k \in \{1, 2\}$,
	\begin{equation}\label{equFCFMajorityAkBkBound}
	\mu(A_k \cup B_k) \leq \frac{\left\lceil\frac{m_i}{2}\right\rceil}{m}.
	\end{equation}
	(This follows since both partitions started with the same number of districts and alternately grew one district at a time until reaching $m_i$ districts, so the maximum number of districts either partition could have at the end is $\left\lceil\frac{m_i}{2}\right\rceil$.) Let $C$ be the remaining part of the interval,
	$$C := \overline{[0, 1] \setminus (A_1 \cup A_2 \cup B_1 \cup B_2)}.$$
	
	There are a few different sub-cases to consider, depending on the advantage of party $i$ in each of these five districts. First suppose that, for some $k_j \in \{1, 2\}$,
	\begin{equation}\label{equFCFMajorityEasyCase1}
	v_i^i(A_{k_j} \cup B_{k_j}) \leq \frac{\mu(A_{k_j} \cup B_{k_j})}{2}.
	\end{equation}
	Then it must be the case that
	\begin{equation}\label{equFCFMajorityEasyCase2}
	v_i^i(\overline{[0, 1] \setminus (A_{k_j} \cup B_{k_j})}) \geq \frac{\mu(\overline{[0, 1] \setminus (A_{k_j} \cup B_{k_j})})}{2},
	\end{equation}
	for otherwise, summing (\ref{equFCFMajorityEasyCase1}) with the negation of (\ref{equFCFMajorityEasyCase2}), we would have that $v_i^i([0, 1]) < \frac12$, contradicting the assumption that $i$ is a majority party. We apply Lemma \ref{lemCutIterated} to divide $\overline{[0, 1] \setminus (A_{k_j} \cup B_{k_j})}$ into $m - \left\lceil\frac{m_i}{2}\right\rceil$ districts of size $\frac1m$. Property (\ref{itmCutIteratedProportional}) of Lemma \ref{lemCutIterated} and inequality (\ref{equFCFMajorityEasyCase2}) imply that party $i$ will win all of these districts (as long as we break ties in favor of $i$). Thus, using these disjoint districts to extend $(P'_{k_j}, T'_{k_j})$ (which is itself an extension of $(P_{k_j}, T_{k_j})$), by Lemma \ref{lemFCFMajoritySufficientCondition}, we have met the geometric target for party $i$.
	
	Now suppose instead that, for all $k \in \{1, 2\}$,
	\begin{equation}\label{equFCFMajorityHardCase1}
	v_i^i(A_k \cup B_k) \geq \frac{\mu(A_k \cup B_k)}{2}.
	\end{equation}
	If, in addition, we have
	$$v_i^i(C) \geq \frac{\mu(C)}{2},$$
	then inequality (\ref{equFCFMajorityEasyCase2}) clearly still holds for either choice of $k_j$, so the same argument goes through. Thus, assume
	\begin{equation}\label{equFCFMajorityHardCase2}
	v_i^i(C) \leq \frac{\mu(C)}{2}.
	\end{equation}
	
	We claim that
	\begin{equation}\label{equFCFMajorityHardCase3}
	\mu(C) \leq \mu(A_1) + \mu(A_2).
	\end{equation}
	Suppose toward a contradiction that (\ref{equFCFMajorityHardCase3}) did not hold. Since all three measures are integer multiples of $\frac1m$, this means that
	\begin{equation}\label{equFCFMajorityHardCase4}
	\mu(C) \geq \mu(A_1) + \mu(A_2) + \frac1m.
	\end{equation}
	We proceed similarly as in the last part of the proof of Lemma \ref{lemFCFMinorityValue}. Letting $Y := B_1 \cup B_2 \cup C$, we have
	\begin{align*}
	\mu(Y) &= \mu(B_1) + \mu(B_2) + \mu(C)\\
	&\geq \mu(A_1) + \mu(A_2) + \mu(B_1) + \mu(B_2) + \frac1m \stext{from inequality (\ref{equFCFMajorityHardCase4})}\\
	&= \frac{m_i}{m} + \frac1m \stext{by the definitions of $(P'_1, T'_1)$ and $(P'_2, T'_2)$}\\
	&= \frac{m_i + 1}{m}\stepcounter{equation}\tag{\theequation}\label{equIVTInvocationContradiction2}.
	\end{align*}
	However,
	$$v_i^i(Y) - \frac{\mu(Y)}{2} \leq 0$$
	from inequality (\ref{equFCFMajorityHardCase2}) and the fact that party $i$ loses all districts in $B_1$ and $B_2$, and
	$$v_i^i([0, 1]) - \frac12 \geq 0$$
	since party $i$ is a majority party. Therefore, by Lemma \ref{lemIVT}, we have
	$$\mu(Y) < \frac{m_i + 1}{m},$$
	contradicting inequality (\ref{equIVTInvocationContradiction2}).
	
	Thus, we have shown that inequality (\ref{equFCFMajorityHardCase3}) holds. It is therefore possible to subdivide $C$ into two districts $C_1$ and $C_2$ such that, for each $k \in \{1, 2\}$,
	\begin{equation}\label{equFCFMajorityHardCase5}
	\mu(C_k) \leq \mu(A_k).
	\end{equation}
	Since $i$ is a majority party, and $A_1$, $B_1$, $C_1$, $A_2$, $B_2$, and $C_2$ form a partition of $[0, 1]$ into districts that only overlap at endpoints,
	\begin{align*}
	0 &\leq v_i^i([0, 1]) - \frac{\mu([0, 1])}{2}\\
	&= \left(v_i^i(A_1 \cup B_1 \cup C_1) - \frac{\mu(A_1 \cup B_1 \cup C_1)}{2}\right) + \left(v_i^i(A_2 \cup B_2 \cup C_2) - \frac{\mu(A_2 \cup B_2 \cup C_2)}{2}\right),
	\end{align*}
	so the two terms in parentheses cannot both be negative. Let $k_i \in \{1, 2\}$ be such that
	\begin{equation*}\label{equFCFMajorityHardCase6}
	v_i^i(A_{k_i} \cup B_{k_i} \cup C_{k_i}) \geq \frac{\mu(A_{k_i} \cup B_{k_i} \cup C_{k_i})}{2}
	\end{equation*}
	and let $k_j \in \{1, 2\}$ be the other index, so $k_i \neq k_j$. As was done in the case where party $i$ was the minority party, we extend $(P'_{k_j}, T'_{k_j})$ (which is itself an extension of $(P_{k_j}, T_{k_j})$) by applying Lemma \ref{lemCutIterated} to $v_i^i$ with
	$$s := \frac{1}{m\mu(A_{k_i} \cup B_{k_i} \cup C_{k_i})}$$
	to cut $\left\lfloor m \mu(A_{k_i} \cup B_{k_i} \cup C_{k_i})\right\rfloor$ districts from $A_{k_i} \cup B_{k_i} \cup C_{k_i}$, which we can ensure are all won by party $i$ by breaking ties in favor of party $i$. Note that these districts clearly have the target size $\frac1m$ from property (\ref{itmCutIteratedMeasure}) of Lemma \ref{lemCutIterated}. The remainder of $[0, 1]$ can be partitioned arbitrarily; denote by $(P, T)$ the resulting $m$-partition of $[0, 1]$. Recall that party $i$ also wins all $m \mu(A_{k_j})$ districts from $A_{k_j}$. Thus, the total number of districts they win is
	\begingroup\allowdisplaybreaks\begin{align*}
	u_i^i(P, T) &\geq \left\lfloor m \mu(A_{k_i} \cup B_{k_i} \cup C_{k_i})\right\rfloor + m \mu(A_{k_j})\\
	&= m \mu(A_{k_i} \cup B_{k_i} \cup C_{k_i}) + m \mu(A_{k_j}) \stext{since $\mu(A_{k_i} \cup B_{k_i} \cup C_{k_i})$ is a multiple of $1/m$}\\
	&= m \left(\mu(A_{k_i}) + \mu(B_{k_i}) + \mu(C_{k_i}) + \mu(A_{k_j})\right)\\
	&\geq m \left(\mu(A_{k_i}) + \mu(B_{k_i}) + \mu(C_{k_i}) + \mu(C_{k_j})\right) \stext{from inequality (\ref{equFCFMajorityHardCase5})}\\
	&= m \left(1 - \mu(A_{k_j}) - \mu(B_{k_j})\right)\\
	&= m - m\mu(A_{k_j} \cup B_{k_j})\\
	&\geq m - \left\lceil\frac{m_i}{2}\right\rceil \stext{from inequality (\ref{equFCFMajorityAkBkBound})}.
	\end{align*}\endgroup
	Hence, by Lemma \ref{lemFCFMajoritySufficientCondition}, the geometric target of party $i$ is satisfied. \qed

\section{Empirical results omitted from Section~\ref{exp:disagree}}\label{appEmpirical}

In Section~\ref{exp:disagree} we report the effect of enforcing the geometric target constraint on competitiveness and compactness in Texas. Here we report the full results for the range of deviations considered.

\begin{figure}[t]
	\begin{subfigure}[b]{0.45\textwidth}
		\centering
		\includegraphics[trim=0 0 0 0.5cm, clip, scale=.4]{{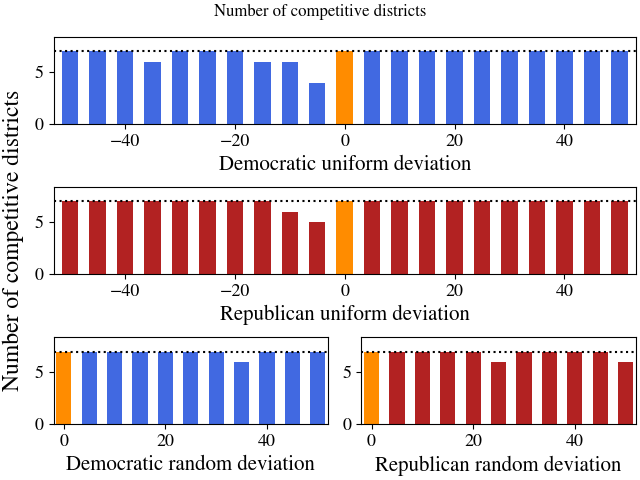}} 
		\caption*{Georgia}
	\end{subfigure} \hspace{5mm}
	\begin{subfigure}[b]{0.45\textwidth}
		\centering
		\includegraphics[trim=0 0 0 0.5cm, clip, scale=.4]{{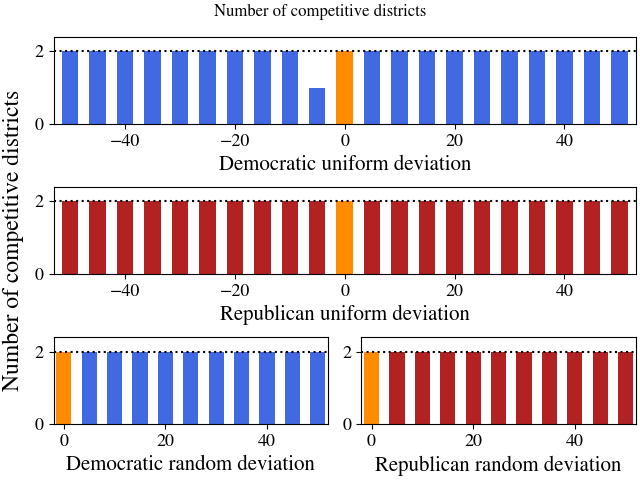}}
		\caption*{Massachusetts}
	\end{subfigure}\vspace{3mm}
	\begin{subfigure}[b]{0.45\textwidth}
		\centering
		\includegraphics[trim=0 0 0 0.5cm, clip, scale=.4]{{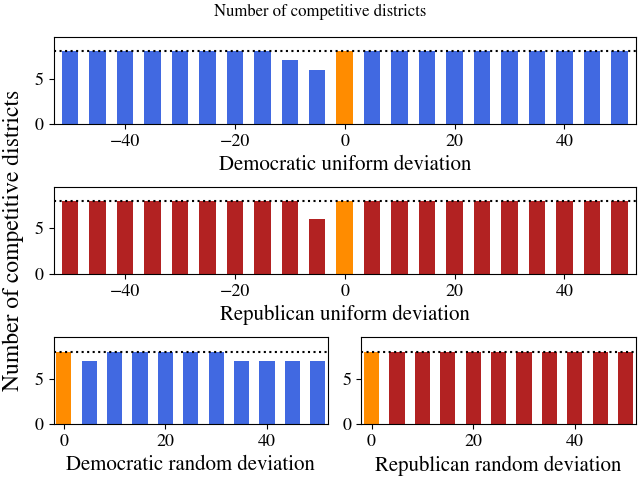}}
		\caption*{North Carolina}
	\end{subfigure}\hspace{5mm}
	\begin{subfigure}[b]{0.45\textwidth}
		\centering
		\includegraphics[trim=0 0 0 0.5cm, clip, scale=.4]{{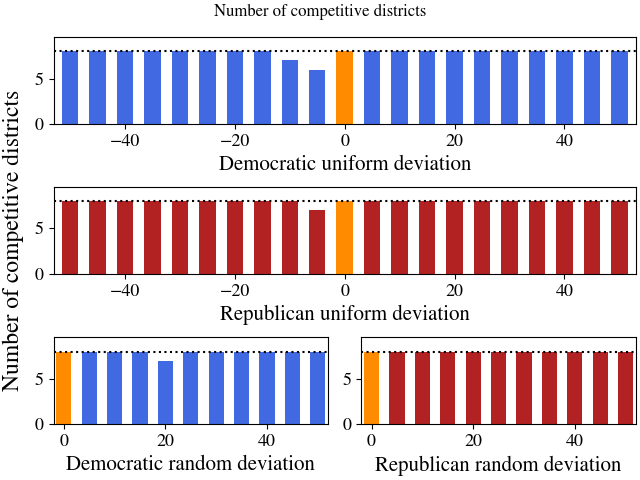}}
		\caption*{Pennsylvania}
	\end{subfigure}\vspace{3mm}
	\begin{subfigure}[b]{0.45\textwidth}
		\centering
		\includegraphics[trim=0 0 0 0.5cm, clip, scale=.4]{{Media/exp/TX_b1000_comp.png}}
		\caption*{Texas}
	\end{subfigure}\hspace{5mm}
	\begin{subfigure}[b]{0.45\textwidth}
		\centering
		\includegraphics[trim=0 0 0 0.5cm, clip, scale=.4]{{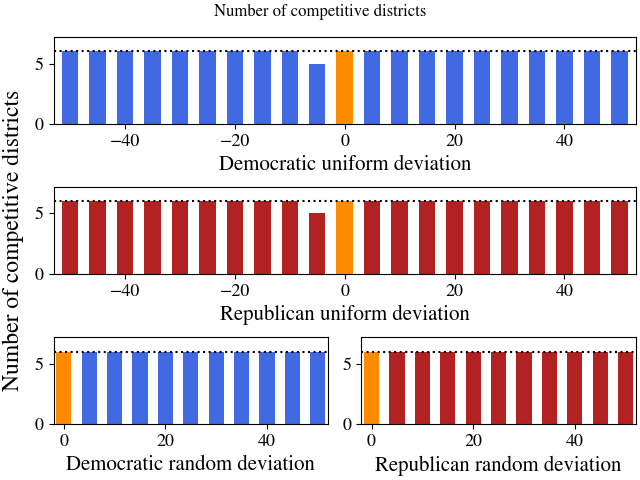}}
		\caption*{Virginia}
	\end{subfigure}
	
	\caption{The largest number of competitive districts among GT partitions compared to the maximum observed (black dotted line) for each of the deviations considered.  The color of the bar  represents which party deviates. The golden bar reports the case when neither party deviates. 
	}
	\label{fig:dev:comp}
\end{figure}

\begin{figure}[t]
	\begin{subfigure}[b]{0.45\textwidth}
		\centering
		\includegraphics[trim=0 0 0 0.5cm, clip, scale=.4]{{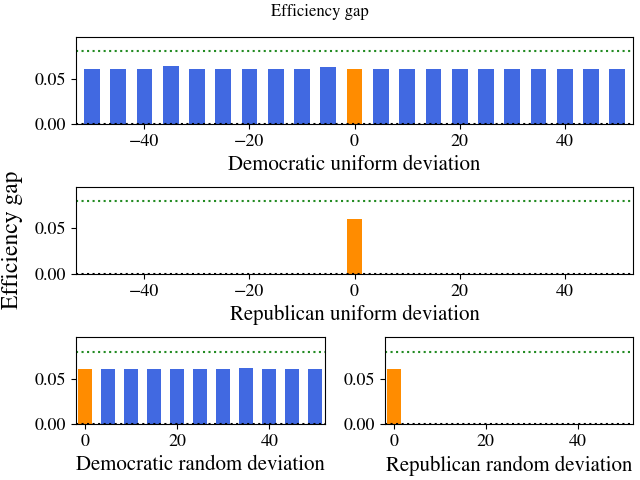}} 
		\caption*{Georgia}
	\end{subfigure}\hspace{5mm}
	\begin{subfigure}[b]{0.45\textwidth}
		\centering
		\includegraphics[trim=0 0 0 0.5cm, clip, scale=.4]{{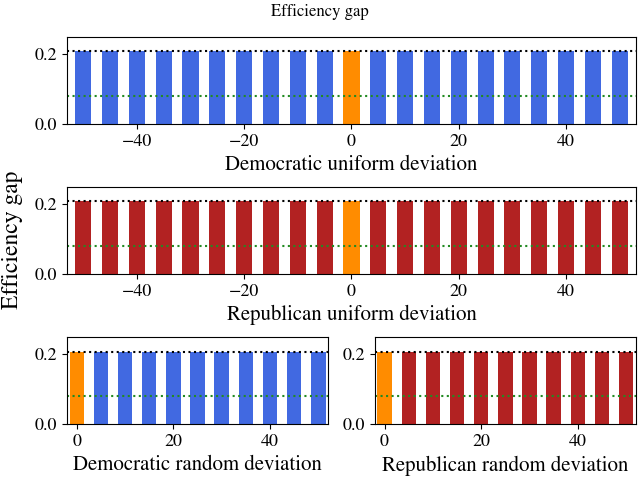}}
		\caption*{Massachusetts}
	\end{subfigure}\vspace{3mm}
	\begin{subfigure}[b]{0.45\textwidth}
		\centering
		\includegraphics[trim=0 0 0 0.5cm, clip, scale=.4]{{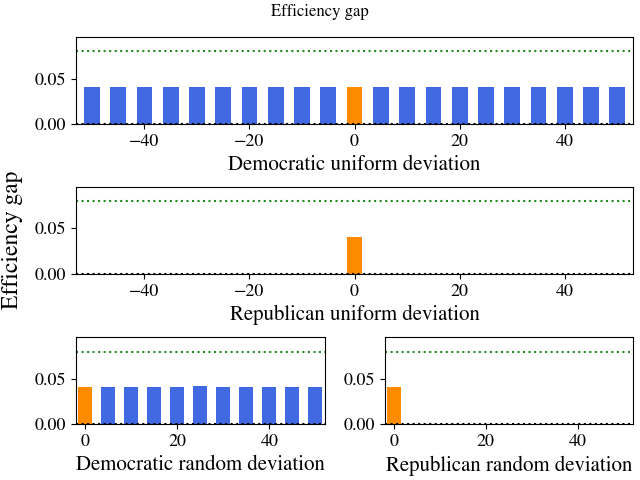}}
		\caption*{North Carolina}
	\end{subfigure}\hspace{5mm}
	\begin{subfigure}[b]{0.45\textwidth}
		\centering
		\includegraphics[trim=0 0 0 0.5cm, clip, scale=.4]{{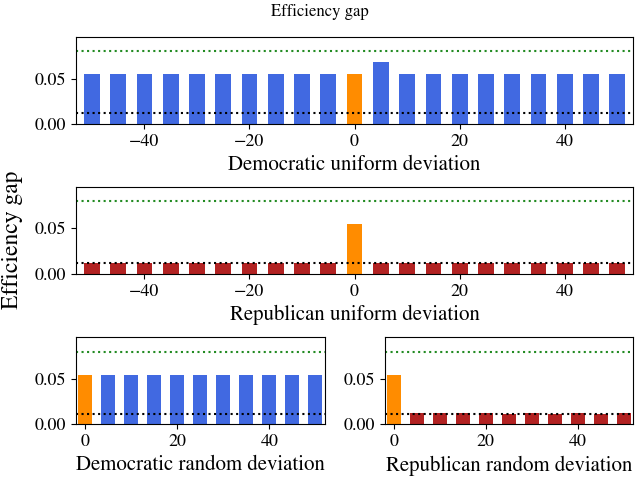}}
		\caption*{Pennsylvania}
	\end{subfigure}\vspace{3mm}
	\begin{subfigure}[b]{0.45\textwidth}
		\centering
		\includegraphics[trim=0 0 0 0.5cm, clip, scale=.4]{{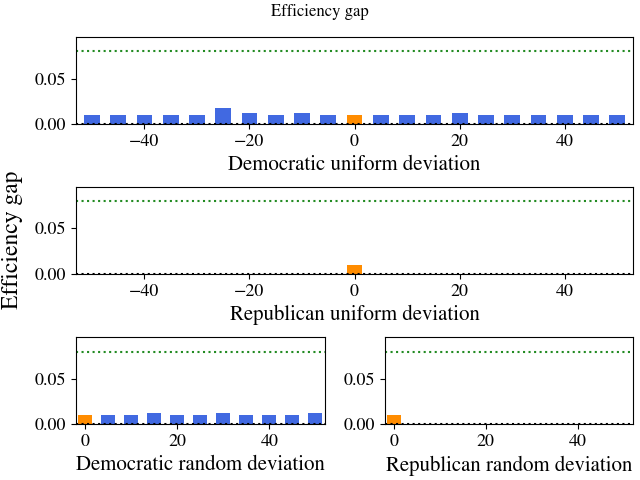}}
		\caption*{Texas}
	\end{subfigure}\hspace{5mm}
	\begin{subfigure}[b]{0.45\textwidth}
		\centering
		\includegraphics[trim=0 0 0 0.5cm, clip, scale=.4]{{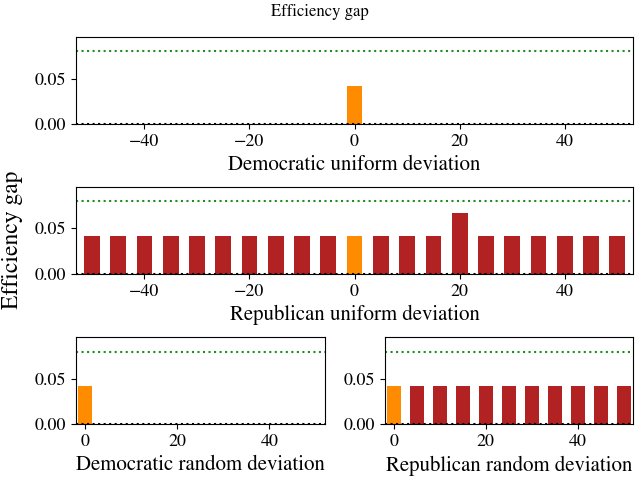}}
		\caption*{Virginia}
	\end{subfigure}
	
	\caption{The smallest absolute efficiency gap among GT partitions compared to the best observed efficiency gap (black dotted line), and a threshold of 8\% (green dotted line).  The color of the bar  represents which party deviates. The golden bar reports the case when neither party deviates.  
	}
	\label{fig:dev:eff}
\end{figure}
\begin{figure}[t]
	\begin{subfigure}[b]{0.45\textwidth}
		\centering
		\includegraphics[trim=0 0 0 0.5cm, clip, scale=.4]{{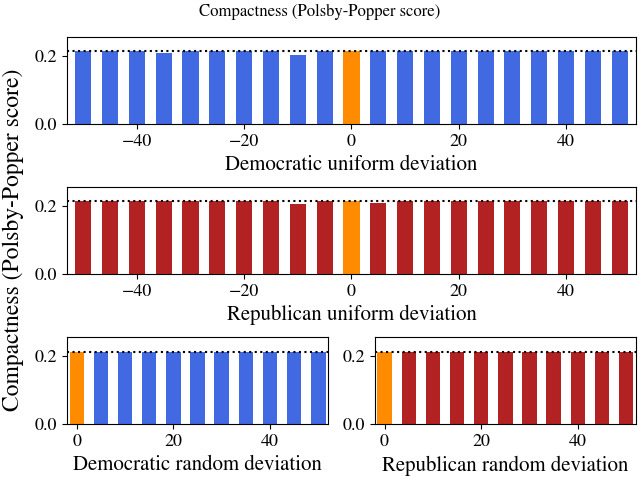}} 
		\caption*{Georgia}
	\end{subfigure}\hspace{5mm}
	\begin{subfigure}[b]{0.45\textwidth}
		\centering
		\includegraphics[trim=0 0 0 0.5cm, clip, scale=.4]{{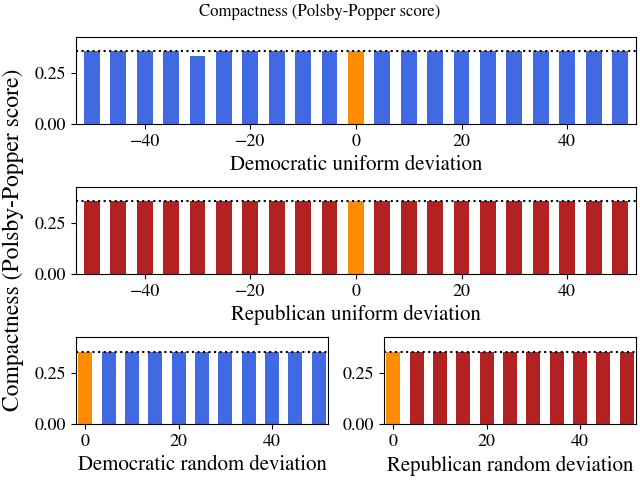}}
		\caption*{Massachusetts}
	\end{subfigure}\vspace{3mm}
	\begin{subfigure}[b]{0.45\textwidth}
		\centering
		\includegraphics[trim=0 0 0 0.5cm, clip, scale=.4]{{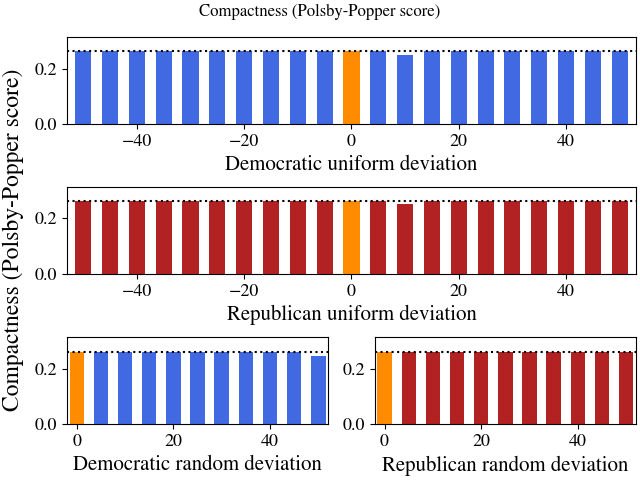}}
		\caption*{North Carolina}
	\end{subfigure}\hspace{5mm}
	\begin{subfigure}[b]{0.45\textwidth}
		\centering
		\includegraphics[trim=0 0 0 0.5cm, clip, scale=.4]{{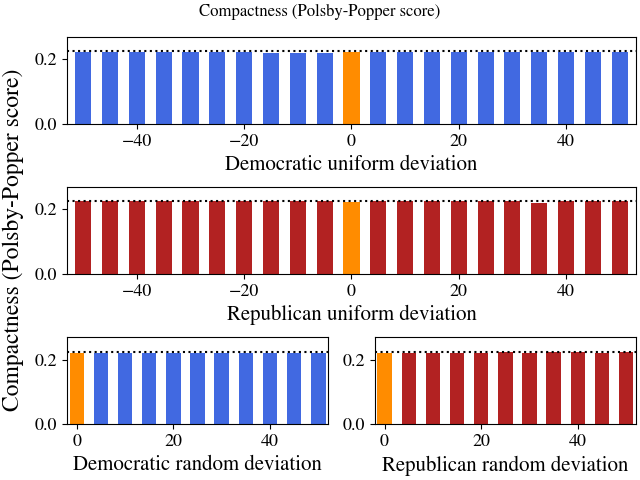}}
		\caption*{Pennsylvania}
	\end{subfigure} \vspace{3mm}
	\begin{subfigure}[b]{0.45\textwidth}
		\centering
		\includegraphics[trim=0 0 0 0.5cm, clip, scale=.4]{{Media/exp/TX_b1000_polsby.png}}
		\caption*{Texas}
	\end{subfigure}\hspace{5mm}
	\begin{subfigure}[b]{0.45\textwidth}
		\centering
		\includegraphics[trim=0 0 0 0.5cm, clip, scale=.4]{{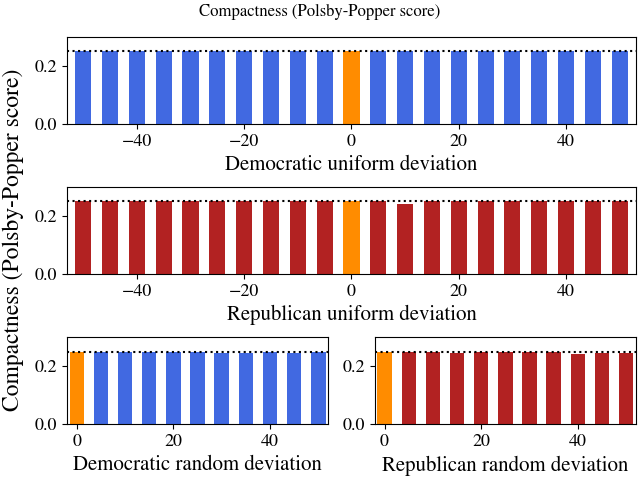}}
		\caption*{Virginia}
	\end{subfigure}
	
	\caption{The most compact GT partitions compared to the best Polsby-Popper score observed (black dotted line) for each of the deviations studied.  The color of the bar  represents which party deviates. The golden bar reports the case when neither party deviates.  
	}
	\label{fig:dev:pp}
\end{figure}

\end{document}